%% file: main.tex
\pdfoutput=1
\documentclass[twocolumn]{aastex631}
\usepackage[T1]{fontenc}
\let\tablenum\relax
\usepackage{siunitx} 
\usepackage{savesym}
\restoresymbol{SIX}{tablenum}
\usepackage{wrapfig}
\graphicspath{{}} %

\newcommand{\tess}{\textit{TESS}}
\newcommand{\tnt}{\textsc{Dynamite}}
\newcommand{\Nsysnewpl}{24}
\newcommand{\Nnewpl}{31}
\newcommand{\Nnewtrpl}{25}

\shortauthors{Turtelboom et al.}

\begin{document}

\submitjournal{ApJ}

\title{Searching for Additional Planets in \textit{TESS} Multi-Planet Systems: Testing Empirical Models Based on \textit{Kepler} Data}
\shorttitle{Testing Empirical Kepler Models of Planetary System Architectures}

\correspondingauthor{Emma V. Turtelboom} \email{eturtelboom@berkeley.edu}

\input{authors}

\begin{abstract}
Multi-planet system architectures are frequently used to constrain possible formation and evolutionary pathways of observed exoplanets. Therefore, understanding the predictive and descriptive power of empirical exoplanetary system models is critical to understanding their formation histories. We analyze 52 \tess{} multi-planet systems previously studied using \tnt{} \citep{dyn2020}, who used \tess{} data alongside empirical models based on \textit{Kepler} planets to predict additional planets in each system. We analyze additional \tess{} data to search for these predicted planets. We thereby evaluate the degree to which these models can be used to predict planets in \tess{} multi-planet systems. Specifically, we study whether a period ratio method or clustered period model is more predictive. We find that the period ratio model predictions are most consistent with the planets discovered since 2020, accounting for detection sensitivity. However, neither model is highly predictive, highlighting the need for additional data and more nuanced models to describe the full population. Improved eccentricity and dynamical stability prescriptions incorporated into \tnt{} provide a modest improvement in the prediction accuracy. We also find that the current sample of 183 \tess{} multi-planet systems are highly dynamically packed, and appear truncated relative to detection biases. These attributes are consistent with the \textit{Kepler} sample, and suggest an efficient formation process. 
\end{abstract}

\section{Introduction} \label{sec:intro}

A large fraction of exoplanets with orbital periods $\lesssim 100$ d are members of multi-planet systems \citep[e.g.][]{lissauer+2011,sandford+2019}. Of the 500 confirmed planets in the systems discovered by the Transiting Exoplanet Survey Satellite (\tess{}), 243 are in 97 multi-planet systems\footnote{NASA Exoplanet Archive, \url{https://exoplanetarchive.ipac.caltech.edu/}, accessed 12 June 2024.}. This sample builds upon the extensive population of multi-planet systems found by the \textit{Kepler} mission, which discovered 525 multi-planet systems containing over 1300 confirmed planets. Extensive analyses of the \textit{Kepler} multi-planet systems \citep[e.g.][]{xie+2016} have revealed demographic features such as the excess of planet pairs wide of resonance \citep{steffen+2015, fabrycky+2014}, and the tendency towards correlated radii and masses of small ($<4R_\oplus$) planets in multi-planet systems \citep{millholland+2017, weiss+2018}. 

Planetary system architectures and demographics encapsulate the outcomes of a diverse set of formation and evolutionary processes, wherein planets dynamically influence each other throughout their lifetimes \citep[e.g.][]{Chatterjee2008, Juric2008, Izidoro2017, Goldberg2022}. Theoretical models of multi-planet systems aim to connect relevant physical processes with the observed population of planets (see \citealt{mordasini18, drazkowska+23} and references therein). Empirical models, on the other hand, seek to characterize the detection biases and observational strategies involved in generating the observed planet population \citep[e.g.][]{he+2019, mulders+2018, macdonald+20}. 

Large-scale transit and radial velocity surveys are currently the largest contributors ($74\%$ and $19\%$, respectively$^{1}$) to the confirmed exoplanet sample. Additionally, complementary methods such as direct imaging, astrometry, and microlensing are probing the sample of planets at wide separations. While all of these detection methods are primarily sensitive to giant planets, the high signal-to-noise ratios of past observations have enabled the detection of thousands of sub-Jovian planets at a large range of orbital periods. Transit and radial velocity surveys preferentially detect planets at short orbital periods. However, direct imaging is particularly sensitive to long-period planets (at 10s of AU), with a bias towards young (and thus, luminous) planets \citep{chauvin23}. The microlensing method is also primarily sensitive to planets at moderately wide separations \citep[$\sim$1-10 AU, e.g.][]{bennett+rhie02, sumi+11, barclay+17}. 

Future missions will use direct imaging and microlensing to probe the outer regions of planetary systems, which are currently under-sampled. The Nancy Grace Roman Space Telescope \citep{spergel+15} is predicted to discover $\sim$1000 bound and $\sim$250 free-floating planets via microlensing \citep{penny+19}, as well as $\sim$100,000 transiting planets \citep{montet+17, wilson+23, tamburo+23}. By detecting both transiting (and thus, likely close-in) and microlensed (at preferentially intermediate separations) planets, Roman will provide a large-scale view of exoplanet systems \citep{montet+17} relative to our current understanding. 

Robust models of the inner regions of planetary systems will be critical to contextualize these observations. However, we must be cautious when using models of exoplanetary systems based on prior surveys which are not generalizable, which may introduce unknown biases in future analyses and cloud our picture of planet demographics.

Several models have been put forth to describe the \textit{Kepler} multi-planet systems \citep[e.g.,][]{mulders+2018, batalha14, zhu+18, yang+20}. These models characterize the detection sensitivity of the \textit{Kepler} mission, and aim to describe the observed exoplanet population and distribution of planetary system architectures. A ``\textit{Kepler} multi'' has come to mean a largely co-planar compact multi-planet system hosting relatively small, roughly uniformly sized planets with similar masses evenly spaced in log-period space. However, this architecture does not describe the full \textit{Kepler} sample of planets, and may not be generalizable to other observed populations. We seek to investigate whether two of the empirical models developed using the \textit{Kepler} sample accurately describe the multi-planet systems discovered by \textit{Kepler}'s successor, the \textit{TESS} mission. 

The empirical models that we study were initially introduced by \citet{mulders+2018} and \citet{he+2019}. In 2020, these models were used to predict an additional planet in the \tess{} multi-planet systems using \tnt{} (the DYNAmical Multi-planet Injection TEster\footnote{\url{https://github.com/jamiedietrich/dynamite}}, \citealt{dyn2020}, henceforth D20). We measure the descriptive power of a model by evaluating how accurately it is able to predict additional planets in a given system. 

This work is structured as follows: in Section \ref{sec:sample}, we discuss the sample of \tess{} multi-planet systems with additional \tnt{} predicted planets, as well as the current sample of \tess{} multi-planet systems. In Section \ref{sec:methods}, we test the model predictions by comparing the planets predicted by these models to the new planets detected with additional \tess{} observations and radial velocity (RV) data since 2020. We then quantify the detection sensitivity for each predicted additional planet using injection-recovery tests and analytical signal-to-noise calculations. In Section \ref{sec:results}, we evaluate which model most accurately predicted the new planets found since 2020, and apply an updated version of \tnt{} to predict additional planets in the current \tess{} sample of multi-planet systems. We discuss our results in Section \ref{sec:discussion}, and conclude in Section \ref{sec:conclusion}.

\section{Data} \label{sec:data}
The \tess{} mission is an all-sky survey searching for transiting exoplanets, and has completed over 4 years of observations since its launch in 2019. Observations are taken in sectors, each covering $24\arcsec$x$96\arcsec$ and lasting two spacecraft orbits ($\sim27$ days). \tess{} has completed its Prime (Cycles 1 and 2) and First Extended (Cycles 3 and 4) missions, and has re-observed stars across multiple sectors. All of the stars discussed in this paper were observed by \tess{} for at least one sector. We devote the majority of this paper to discussing a subset of 52 stars that were first observed prior to February 2020 and then re-observed (albeit not continuously) between February 2020 and March 2024. For readability, we refer to \tess{} observations of our sample up to and including sector 21 as ``initial'' observations, data collected between sectors 22 and 76 inclusive as ``recent'' observations, and all available \tess{} observations of a target up to and including sector 76 as ``full'' observations.

\subsection{Initial \tess{} Observations} 
In 2020, \tess{} had completed 21 sectors of observations, and had detected 1800 \tess{} Objects of Interest (TOIs). D20 investigated the architectures of the 52 \tess{} multiplanet systems that had been discovered at the time. In the Prime mission, targets selected for high cadence monitoring were initially observed at 2-minute cadence, and full-frame images (FFIs) were collected every 30 minutes. D20 used \tess{} data up to and including Sector 21 to calculate posterior period and radius probability distributions for an additional planet in each of the 52 multi-planet systems. The \tess{} data available varied substantially between targets; 16 stars were only observed for 1 sector, while 8 stars had 10 or more sectors of data. Most of the observations (for 33/52 targets) included small data gaps of 1 or 2 sectors ($\sim 30 - 60$ days), but 5 targets were observed at times 3 or more sectors apart. 

\subsection{Recent \tess{} Observations}
We define the recent \tess{} observations as spanning sectors 22 to 76, which include the end of the Prime mission, the First Extended mission, and the start of the Second Extended mission. \tess{}'s First Extended mission covered sectors 27-55 (Cycles 3 and 4), and lasted approximately two years. For these sectors, an even shorter exposure time of 20 seconds was introduced alongside the 2-minute cadence, and the FFI exposure time was reduced to 10 minutes. In the Second Extended mission (Cycles 5-7, sectors 56-96), the FFI exposure time was further reduced to 200 seconds. As of August 1 2024, \tess{} is conducting observations of sector 81. As in the initial \tess{} observations, the data coverage was heterogeneous; 5 targets were only observed for 1 sector out of 54, while 15 of the 52 targets were observed for more than 15 sectors between sectors 22 and 76 (inclusive). 

\section{Sample} \label{sec:sample}
The goals of this paper are to assess the extent to which empirical models can be used to predict the presence of additional planets in multiplanet systems, and to analyze the architectures of these systems. Accordingly, our primary target sample is a set of planetary systems that were first observed prior to 2020 and then re-observed by \tess{} in subsequent sectors. D20 analyzed the initial \tess{} photometry to make predictions about additional planets. 

Specifically, they considered the set of 52 \tess{} targets with multiple detected planets or planet candidates and made predictions about the periods and radii of possible additional planets in those systems. D20 introduced and made use of the software package \tnt{}. This framework returns posterior probability distributions for planet radius and orbital period under two empirical models: a clustered period model, and a period ratio model. The period ratio model (henceforth PRM) suggests that planet pairs in a multi-planet system have similar spacing in log-period space, as in \citet{mulders+2018}. The clustered period model (henceforth PCM) suggests that planets in multi-planet systems are found in clusters centered on periods that are determined by a system's period distribution \citep{he+2019}. 

 \tnt{} calculates the likelihood of finding another planet in a multi-planet system by integrating over the probability density functions of planet inclination, period, and radius given the known planets in the system. A key assumption made in this formalism is that these three probability density functions are separable. This is likely untrue, as systems with low mutual inclinations typically host smaller planets with shorter orbital periods \citep{Dawson2016}, among other apparent correlations between planet parameters \citep[e.g.][]{millholland+2017, Carrera2018}. However, in the absence of well-defined correlation functions, independent probability density functions for orbital period, radius, and inclination are used in \tnt{}. D20 used the period probability density functions associated with the PCM and PRM to predict an additional planet in each of the 52 \tess{} multi-planet systems known in 2020, generating two sets of posterior probability distributions.

In this paper, we first use more recently obtained \tess{} photometry to assess the performance of the model predictions applied by D20. We discuss the sample of targets used in D20 in Section~\ref{sec:dyn20sample}. Next, we broaden our analysis to the current sample of systems with multiple transiting planets or planet candidates observed by \tess{}. We discuss this larger sample in Section \ref{sec:dyn23sample}.

\subsection{\tess{} Multi-Planet Systems Observed Prior to 2020 \& Subsequently Re-Observed} \label{sec:dyn20sample}

D20 created their sample by selecting stars in the ExoFOP-\tess{} archive which hosted more than one TOI. They excluded four systems which hosted planets $>5R_\oplus$, as the period distributions implemented in \tnt{} are not representative of giant planets \citep[e.g.][]{dong&zhu13}. Their sample included 39 systems hosting 2 TOIs, 11 systems hosting 3 TOIs, and 2 systems hosting 4 TOIs. We refer to this sample throughout the paper as the ``2020 sample.'' They only included transiting planets and planet candidates as inputs to their models as \tnt{} returns posterior distributions for planet radius. The posterior distributions returned by \tnt{} under the clustered period model strictly peaked at shorter periods than those under the period ratio model ($P_{median, PCM} = 2.7$ days, $P_{median, PRM} =$ 20.3 days).

\subsubsection{Insights From Additional Observations}

\begin{figure}
    \includegraphics[width=\columnwidth]{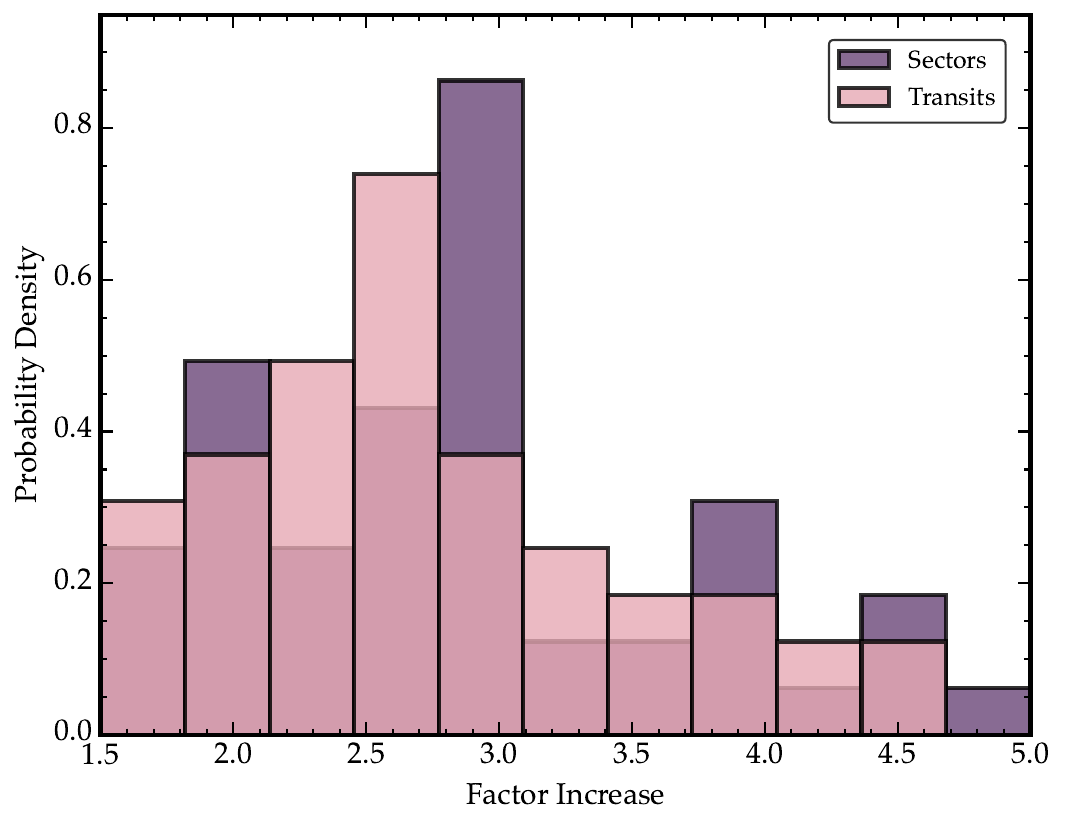}
    \caption{Histogram of the ratio of the total number of \tess{} sectors (up to and including Sector 76) to the number of sectors available in 2020 (prior to Sector 21, in purple) for each system in our sample. We also plot the histogram of the ratio of the number of transits in the currently available data to the number of transits in the pre-sector 21 \tess{} light curves (pink). We calculate this ratio as the mean of transit ratio for each planet candidate in a system. All 52 systems have substantially increased photometric baselines.}
    \label{fig:hist}
\end{figure}

All of the 52 systems in the 2020 sample were re-observed for between 1 and 25 additional sectors and a median of 5 additional sectors in recent \tess{} observations. Fig. \ref{fig:hist} shows that both the number of sectors and the number of transits (assuming linear ephemerides) observed for each target has increased by at least a factor of 2 for the majority of the sample. The candidates with the largest increase in \tess{} data now have six times more sectors of data and 4.5 times more transits than in the initial \tess{} observations. With this extended baseline, we can not only search for longer-period transiting planets, but also improve our detection sensitivity for short-period transiting planets. Of the original sample, \Nsysnewpl{} systems have had additional planets found and reported on ExoFOP\footnote{ExoFOP, \url{https://exofop.ipac.caltech.edu/tess/}, accessed 11 July 2024}, for a total of \Nnewpl{} new confirmed or candidate planets. These newly discovered planets allow us to evaluate the accuracy of the predictions made in 2020.

The newly discovered (candidate) planets are both transiting and non-transiting, and are strikingly diverse; these planets range from the size of Venus to that of Neptune, and have orbital periods ranging from 1.91 d to over 4000 d. Furthermore, the newly discovered planets and planet candidates are interior, in-between, and exterior to previously known planets. 

Since the publication of D20, follow-up observations have been conducted for many of the 52 systems considered in the paper. By analyzing these follow-up observations and the recent \tess{} data, three of the planet candidates in the 2020 sample have been revealed to be False Alarms. Additionally, the number of planets in another system in the original sample (TOI-1339) has also been the subject of debate. We discuss these four systems below.

\begin{enumerate}
    \item TOI-282, also known as HD 28109, was thought to host four transiting planet candidates in 2020. The planet candidate TOI-282.02 had an orbital period of 31 d, was included in the D20 analysis of this system. However, it has since been found to be a False Alarm. The two outermost planets in this system are near the 3:2 MMR and all three confirmed planets exhibit transit timing variations (TTVs) of up to 60 minutes. These TTVs induced the signal initially labelled as TOI-282.02. We include the three confirmed transiting planets ((HD 28109 b: $22.9 d$, $2.199^{+0.098}_{-0.10} R_\oplus$, $18.5^{+9.1}_{-7.6} M_\oplus$, HD 28109 c: $56.0 d$, $4.23 \pm 0.11 R_\oplus$, $7.94^{+4.23}_{-3.05} M_\oplus$, and HD 28109 d: $84.3 d$, $3.25 \pm 0.11 R_\oplus$, $5.69^{+2.74}_{-2.11} M_\oplus$, \citealt{dransfield+2022}) in all analyses in this work (Section 4.2 onwards).
    \item TOI-1277 is a K dwarf that had two planet candidates at the time of publication of D20. TOI-1277.02, with a period of 14.9 days, is still classified as an active planet candidate. TOI-1277.01, on the other hand, only had 1 associated transit-like event, and has been retired as a False Alarm. Therefore, this system currently hosts only one planet or planet candidate, and we exclude it from further analysis (see Section \ref{sec:dyn23sample}). 
    \item TOI-1449 is a M dwarf which was previously thought to host two planet candidates. Similarly to TOI-1277, one of the planet candidates (TOI-1449.01) only had one transit-like event, and has been retired as a False Alarm. There is currently with one active planet candidate at 2.3 d, TOI-1449.02. We therefore exclude this system from the updated sample list (see Section \ref{sec:dyn23sample}).
    \item TOI-1339, also known as HD 191939, is a high multiplicity system, and there is some discussion in the literature about the number of planets in the system \citep{lubin+24, badenas-agusti+2020, orell-miquel+2023}. At the time of D20, the system was believed to host 3 planet candidates, with periods of $\sim8, \sim25$ and $\sim28$ d. As of 12 June 2024, the system hosts three confirmed planets with updated periods of 8.9 d, 28.6 d, and 38 d \citep[HD 191939 b: 8.9 d, $3.41 \pm 0.08 R_\oplus$, $10.0 \pm 0.7 M_\oplus$, HD 191939 c: 28.6 d, $3.20 \pm 0.08 R_\oplus$, $8.0 \pm 1.0 M_\oplus$, HD 191939 d: 38.4 d, $3.00 \pm 0.07 R_\oplus$, $2.8 \pm 0.6 M_\oplus$, ][]{badenas-agusti+2020}. The inaccuracies of the earlier period measurements were likely due to the fact that the host star fell outside of the CCD’s science image area in sector 14 \citep{badenas-agusti+2020}, which was not flagged in the initial \tess{} Quick-Look Pipeline planet search \citep{qlp}. \\
    The HD 191939 system also includes non-transiting planets.\citet{lubin+22} and \citep{lubin+24} reported 2 additional non-transiting planets in the system (HD 191939 e: 101.7 $\pm$ 0.1 d, $Msin(i) = 114.1 \pm 3.1 M_\oplus$, HD 191939 f: 2898 $\pm$ 152 d, $Msin(i) = 2.88 \pm 0.26 M_{Jup}$). \citet{orell-miquel+2023} found a third non-transiting planet in the Habitable Zone of this system (HD 191939 g: $284^{+10}_{-8} d$, $Msin(i) = 13.5 \pm 2.0 M_\oplus$), for a total of six confirmed planets. We incorporate the six confirmed planets HD 191939 b-g in our analysis.
\end{enumerate}

\subsection{\tess{} Multi-Planet Systems in 2024} \label{sec:dyn23sample}
As well as increasing the baseline of the 52 systems in the 2020 sample, \tess{}'s continued observations have revealed additional multi-planet systems. As of 26 Feb 2024, there are 183 TOIs with at least 2 planet candidates (PCs), confirmed planets (CPs), ambiguous planet candidates (APCs), or known planets (KPs)$^{3}$. We note that only 50 of the 52 TOIs analysed in D20 are included in the current sample, as two systems in the original sample currently have only one PC, APC, CP, or KP (TOI-1277 and TOI-1449). We refer to this sample of 183 multi-planet systems as the ``2024 sample''.

\section{Methods} \label{sec:methods}
In this section, we describe the methods used to search the full \tess{} observations for the predicted additional planets in each system in the 2020 sample. We used both an empirical (see Section \ref{sec:injrec}) and analytical (see Section \ref{sec:analytical}) approach to quantify the detection sensitivity.

\subsection{Planet Search} \label{sec:transitfit}
We developed a planet search pipeline in order to search for transits (injected or real) in the light curves of targets in our sample. Our pipeline considers a combination of light curves produced by the \textit{TESS} Science Processing Operations Center \citep[SPOC, ][]{spoc} and the \tess{}-\textit{Gaia} Light Curve \citep[TGLC,][]{han+23} pipelines. For each sector of observations, we selected the 120-second SPOC cadence light curve if available. If this data product was not available for a given sector, we used the light curve with the highest cadence (shortest exposure time), in order to maximise the signal-to-noise ratio of transit signals in the data. For some sectors, targets were only observed in the \tess{} Full-Frame Images (FFIs). If only FFI-derived light curves are available, we selected the TGLC data products, which make use of the \textit{Gaia} DR3 data as position priors for individual targets\citep{han+23}. 

We first flattened the light curves using the \texttt{wotan} Python package \citep{wotan}. This procedure uses a B-spline Huber regression to fit variability, with the number of knots set by the window function. We set the window function to 0.5 days, in order to capture the typical time scales of photometric variability in the light curves in our sample. However, this approach limited our sensitivity to detecting transits with durations on the order of 0.4 days or longer, as these features were often flattened out during this step. However, for the bulk of stars in our sample, transit durations of 0.4 days correspond to orbital periods $\gtrsim10$x larger than those of the predicted additional planets. We removed outliers that lie more than 5 standard deviations above the median flux. We did not remove outliers below the median flux in order to avoid removing data points in transits. We then used a Box-Least Squares (BLS) periodogram \citep{bls} to search for periodic transit-like features in the light curves. We used a period grid between 0.5 and 730 days (as in D20) and a duration grid with 50 values between 0.02 and 0.4 days (0.5 - 9.6 hours) when calculating the BLS periodogram. We optimized the period grid spacing in order to limit the periodogram's size to 10,000 points for computational tractability. We searched each light curve 7 times, masking transits (or transit-like events) associated with the strongest peak in the BLS periodogram after each iteration. 

For each candidate period found by these searches, we phase-folded the light curve using the period and epoch associated with the strongest peak in the BLS periodogram in each iteration of the search. We then calculated the mean number of data points between phases of -0.05 and 0.05 of the phase-folded light curve. We discarded search results where this number was more than twice that outside the phase range surrounding the transit. This indicated that the periodic signal identified by this iteration of BLS was a periodic data gap or ramp feature sometimes seen near the start of \tess{} sectors, and was not a transit.

\begin{figure}
    \includegraphics[width=\columnwidth]{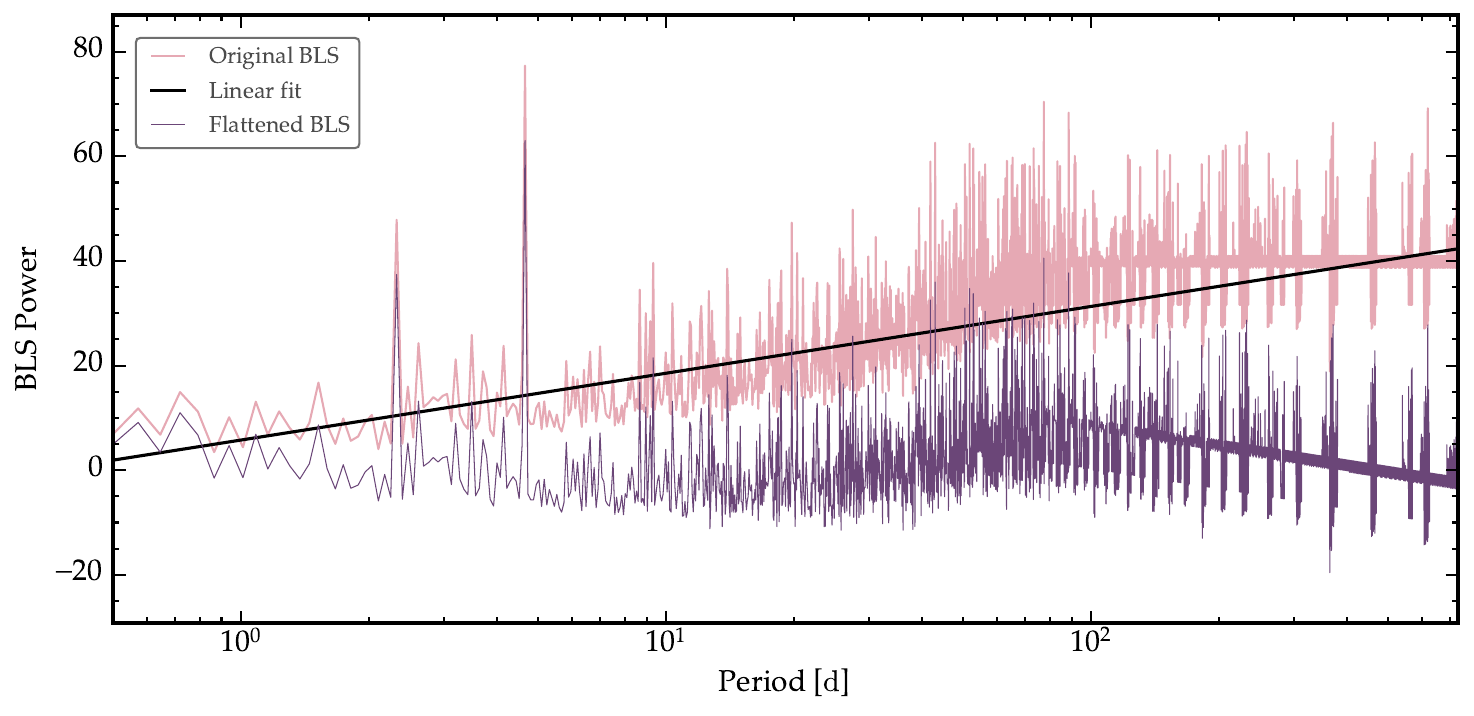}
    \caption{Box-Least Squares periodogram (BLS, pink) of the light curve of  TOI-125 (observed in cycles 1, 3 and 5, with gaps of $\sim$ 2 years between each data segment). Quasi-periodic, heteroskedastic structure is present at periods beyond $\sim$ 50 days. We fit a weighted linear function (black) to the original periodogram (pink) and return a flattened BLS periodogram (purple). Thereby, we remove the general upwards trend with increasing orbital period and thus minimize the impact of the long-period structure on the recovery of injected planets.}
    \label{fig:longperiodbls}
\end{figure}

We observed substantial quasi-periodic, heteroskedastic structure at long periods in the BLS periodograms of several light curves, which we attribute to several-month data gaps between \tess{} observations of a target . For example, TOI-125 (TIC 522368076) has three sectors of \tess{} data, but the system was observed for 1 sector at a time with 2~year gaps between sectors. The BLS periodogram of this light curve is shown in Fig. \ref{fig:longperiodbls}, and exhibits a ramp effect up to $\sim100$ d, after which the periodogram displays quasi-periodic structures. Nine of the 183 systems had predicted additional planets beyond 100 d. In these cases, the peak in BLS periodograms at the predicted period was frequently lost in the long-period structure, which reduced the recovery fraction for these signals. For the 95\% of systems with predicted additional planets within 100 d, the BLS peaks associated with the predicted period were often swamped by this long-period structure. As a result, signals corresponding to the additional predicted planet were not recovered even if the associated BLS peaks were locally maximal. 

The long-period structure observed in the BLS periodograms is likely due to the pattern of \tess{} pointings across cycles. The spacecraft has mostly alternated annually between observing the Northern and Southern hemispheres, with some sectors observing the ecliptic. Previous works have used a variety of models to remove long-term trends from BLS periodograms. These include subtracting the median-filtered periodogram \citep{ofir14}, and quadratic smoothing using 20 equally-spaced spline knots \citep{gondhalekar+23}. In order to mitigate the impact of these features, we fit and subtracted a weighted linear function to the BLS periodogram. The function weighted periodogram values at periods $>100$ d 100 times more strongly than those at shorter periods. This method suppressed the strong, unphysical structure at longer periods in the periodogram, and thus increased the detection sensitivity for planets at all orbital periods.

We also discarded signals which were found multiple times by our algorithm. In each iteration of the planet search, we mask potential `transit events' using the period and epoch associated with the highest power peak in the periodogram. As such, if a candidate period is returned several times, it is likely attributable to the periodic data gaps created by transit masks in previous iterations. Therefore, if a candidate period is returned more than two times by the BLS search, we restrict the period space being searched to be shorter than this period for the remainder of the BLS searches.

\subsection{Transit Injection \& Recovery} \label{sec:injrec}
In order to quantify our detection sensitivity to the predicted additional planets, we performed transit injection and recovery tests. For each target, we injected a \texttt{batman} \citep{batman} transit model into the light curve, after masking the transit events associated with known TOIs in each system. We set the injected planet period and radius using the peak of the respective D20 posterior probability distributions. We drew a transit epoch for the injected planet from a uniform distribution spanning the target's observational baseline, and an impact parameter from a uniform distribution between 0 and 1. We then flattened the light curve and performed an uninformed search for periodic transit-like features as described in Section \ref{sec:transitfit}. Additionally, we considered the 10 strongest peaks in the BLS periodogram in each of the 7 iterations of the planet search, in order to further mitigate the impact of strong non-physical peaks at long periods due to the window function of observations. For each of the 10 strongest peaks in the periodogram, we checked whether the associated period was consistent with the injected period, and whether there was data at phase of 0 when the light curve was folded at this orbital period and associated transit midpoint. 

We consider an injected signal recovered if it has a reported period within 10\% of the injected period (or an integer multiple up to 4 of the injected period). We consider the transit epoch recovered if it falls within 0.1 phase of the injected transit midpoint. We also consider the radius of the planet recovered if it is consistent with the predicted radius to 1$\sigma$ using the errors associated with the predicted radius distribution. We only consider a planet recovered if both the period and transit midpoint are recovered, and if the Signal Detection Efficiency is greater than 7.3 (as is required for detection by the \tess{} SPOC pipeline). We perform 100 injection/recovery tests on each target in the sample, and show the fraction of injected planets recovered in Fig. \ref{fig:injrec_results2020}. 

\subsection{Analytical Detection Sensitivity} \label{sec:analytical}

We also used analytical calculations to evaluate the detectability of each predicted planet. For each target, we generated a grid of planet radius ($R_p$) vs. orbital period ($P$) in the following ranges: $0.5-5~R_\oplus$, $0.5-730~d$. These ranges are consistent with the parameter space explored by D20. For each (P, $R_p$) combination, we calculated the signal-to-noise ratio associated with such a transit, 
\begin{equation}\label{eq_snr}
    SNR = \frac{(R_p/R_*)^2}{CDPP_{\rm{eff}}} \sqrt{\frac{t_{obs}f_0}{P}}
\end{equation}
where $R_*$ is stellar radius, $CDPP_{\rm eff}$ is the effective Combined Differential Photometric Precision, $t_{obs}$ is the time span of the \tess{} observations of a target, and $f_0$ is the duty cycle of the light curve (i.e. the fraction of the baseline during which the target was observed). In order to calculate $CDPP_{\rm eff}$, we used the \texttt{lightkurve.estimate\_cdpp} function, which calculates a quasi-CDPP for a given transit duration and light curve \citep{aigrain+15}. We note that $CDPP_{\rm eff}$ depends on the cadence of the light curve, with longer exposure times (i.e., lower cadence) associated with higher $CDPP_{\rm eff}$ and thus lower SNR for a given set of planet parameters. For light curves made up of sectors with different cadences, we calculate $CDPP_{\rm eff}$ for a putative planet signal for each cadence, using only the sectors with that cadence. We then compute the median $CDPP_{\rm eff}$ across all cadences, and use this value to calculate the SNR. 

We calculated transit durations using the planet period, radius, and inclination at the maximum of the respective posterior probability distributions returned by \tnt{}. We drew eccentricity samples from a truncated Rayleigh distribution with a mode of 0.0355 (as per \citealt{Mills+2019}), and drew inclinations from a uniform distribution of cos(i) ranging from 0 to $cos(i_{\rm{min},\rm{transit}})$, where $i_{\rm{min},\rm{transit}}$ is the minimum inclination possible for a putative planet to still transit the host star. We find that all of the putative additional planets in both models would have calculated SNR $> 7.3$, and thus would pass the detection threshold of the \tess{} mission. This is contrary to the recovery fractions found in Section \ref{sec:injrec}; the 86th percentile of the recovery fraction across our sample is $56\%$ for the PRM and $44\%$ for the PCM. This discrepancy highlights the impact of drawing transit midpoints and impact parameters for the injected planets, which can lead to fewer and shallower transits, respectively. Furthermore, in some cases, the injected flux will not contain any transits due to the placement of injected transits in data gaps. The injection-recovery method also includes a more nuanced treatment of noise. The quasi-CDPP used in the analytical SNR calculations is an approximation of the scatter in a light curve after removing long-term trends; performing injection-recovery tests allowed us to sample the true noise distribution in each light curve.

\section{Results} \label{sec:results}
\subsection{\tnt{} predictions for 2020 sample} 
We set out to determine which model used in D20 to predict additional planets was most accurate, using extended \tess{} observations, injection-recovery tests, and the results of radial velocity studies of these systems. Additional observations of a given system can verify model predictions in several ways: 1) reveal an additional planet with an orbital period corresponding to the peak of the posterior probability distribution, and 2) reveal an additional planet with an orbital period with some other, non-zero posterior probability. We evaluated model performance for case 1 by computing the overlap between the normalized posteriors and probability distributions of new TOIs. In order to do so, we numerically integrated over the intersection of the DYNAMITE posterior set and the newly discovered planet posterior set shaded in grey in Fig. \ref{fig:overlap}). The ideal scenario for case 1 involves a new planet with a tightly constrained period lying at the peak of the \tnt{} posterior period distribution. However, a caveat of the overlap method is that it also favors newly discovered planets with poorly constrained orbital periods (and thus, wide posteriors). For case 2, we measured the difference in period between new TOIs and the peaks in the posterior distributions to evaluate model performance. A representative example is shown in Fig. \ref{fig:overlap}. 

Of the 52 systems analyzed, \Nsysnewpl{} of them have been found to have additional planets, for a total of \Nnewpl{} new planets and planet candidates in these systems. Table \ref{tab:summarydynv1} reports the predicted additional planets and newly detected planets under both models. We find that 8 of the additional planets in the systems have orbital periods that are consistent to $1\sigma$ with the peak period posterior returned by the PRM, and 4 are consistent with the predictions from the PCM to $1 \sigma$. Accordingly, the overlap between the PRM posteriors and the period distributions of the new planets is $\sim1.5$ times that for the PCM posteriors. We note that these results are not mutually exclusive, as the PRM and PCM can predict a similar period range where an additional planet is then found. This is the case for two of the systems (TOI-174 and TOI-178). Of the additional transiting planets found, 15 of them are consistent with the radius predictions to $1 \sigma$ (which are agnostic to the period model), and 9 of these 15 planets also match at least one of the peaks of the period posteriors. These results are also included in Table \ref{tab:summarydynv1}.

\subsection{Updated \tnt{} predictions for 2020 sample}
Since D20, \tnt{} has been expanded with more planet parameter models and a more robust dynamical stability criterion \citep[][]{dietrich+22}. In particular, multiplicity-dependent log-normal distributions for the eccentricity and the mutual inclination were taken from \citet[][]{he+2020}. These distributions empirically matched the observed Kepler population, and were added to \tnt{} to compare with the previous Rayleigh distribution for mutual inclination from \citet[][]{mulders+2018}. We ran the newest version of \tnt{} (henceforth referred to as \tnt{} v3) on the 2020 sample, and found that the new results more accurately predicted the newly discovered planets in the 52 systems in our sample. The resulting predictions from \tnt{} v3 are also reported in Table \ref{tab:summarydynv3}. One exception was in the TOI-797 system, where both the PRM and PCM models in the original version of \tnt{} (henceforth referred to as \tnt{} v1) accurately predicted an additional planet was mostly likely to be found at 2.7 d. However, \tnt{} v3 returned a reduced likelihood of an additional planet at 2.7 d, likely due to the additional dynamical stability constraints implemented in \citet{dietrich+22}.

We found that 13 of the newly discovered planets in the systems matched the PRM period predictions, compared with 8 planets matching the \tnt{} v1 PRM predictions. Three of these 13 matching planets were found iteratively; after one new real planet candidate was discovered in the system, we added that planet candidate to the set of known data with period and radius set to the discovered values and ran \tnt{} v3 again. 11 of the new planets matched the predictions from the PCM (relative to 4 matches found previously), the same 3 of which were found iteratively. The PRM and PCM predictions were more consistent for these systems using this version of \tnt{} (v3) than the previous version (v1), with 8 of the planets found by both orbital period models and 16 of them found by at least one model. Of the \Nnewtrpl{} additional planets that were found to transit, 17 of them were consistent with the radius predictions from \tnt{} v3 to $1 \sigma$. Including the eccentricity distribution and an updated orbital inclination distribution improved the efficacy of the new dynamical stability criterion used to determine the likelihood of additional planets in close pairs with known planets, as dynamical stability and dynamical packing is significantly affected by planet orbital eccentricity \citep[e.g.,][]{obertas+23}. These changes particularly improved the performance of PCM, as the period clustering model tends to pack planets as close together as is dynamically possible.  This was reflected in the calculated overlap between the updated PCM and PRM posteriors and the newly discovered planets. We found that the overlap between the updated PCM posteriors and the newly discovered planets was 1.1 times that for the updated PRM posteriors and newly discovered planets; the two models performed similarly when including a nuanced eccentricity and inclination treatment.

\subsection{\tnt{} v3 predictions for 2024 sample} \label{sec:dyn23}
We also considered the 2024 sample of 183 multi-planet systems observed by \tess{}, as defined in Section \ref{sec:dyn23sample}. We run \tnt{} v3 on this sample, and present our results in Fig. \ref{fig:injrec_results2023} and Tables \ref{tab:tnt2023pcm} and \ref{tab:tnt2023prm}. Of this sample, 136 targets (66\%) have been re-observed since sector 76 or will be re-observed during the planned \tess{} Cycle 7 observations. These stars will receive between 1 and 11 (with a median of 4) additional sectors of data. We find that $\sim80\%$ of predicted additional planets have recovery fractions $<50\%$ with the full \tess{} data (up to and including Sector 76), suggesting that these additional planets may be challenging to detect in the current \tess{} data set. \tess{} observations beyond sector 76 will increase the number of sectors by a median factor of 1.25, and provide a similar boost in the number of observable transits of predicted planets. These additional \tess{} observations will improve our sensitivity to detecting the predicted additional planets in each system. Figures \ref{fig:injrec_results2023} and \ref{fig:rp} show the peaks of the most up-to-date \tnt{} posteriors for the 183 \tess{} multi-planet systems, as well as the recovery fractions for these predicted planets. For the majority of systems (152/183), the PCM posteriors peak at lower periods than the PRM posteriors. Additionally, the PCM posteriors peak interior to the innermost known planet for the majority of systems (125/183), rather than exterior to the outermost known planet (10/183) or in a gap between two known planets (47/183). Under the PRM, the posteriors rarely peak interior to the innermost known planet (26/183), predominantly peak exterior to the outermost known planet (126/183), and peak in a gap for 34/183 systems. 

The recovery fractions for the PRM predicted planets are typically low. Even if the outer planets predicted under the PRM were to exist, we would not be sensitive to detecting them in the \tess{} observations up to and including Sector 76. Furthermore, the median transit probability of predicted additional planets under either model is $\sim 0.5$ (based on the geometric transit probability associated with the planets' predicted inclinations, listed in Tables \ref{tab:tnt2023pcm} and \ref{tab:tnt2023prm}). Therefore, there is an even chance that if the predicted outer PRM planets were to exist, they would not transit and therefore be undetectable by \tess{}, regardless of their radii. Additional observations through existing and future radial velocity surveys will reveal whether some of these predicted planets exist but do not transit their host stars.

On the other hand, the majority of the additional planets predicted using the PCM have short orbital periods and high recovery fractions (see Fig. \ref{fig:injrec_results2023}). As such, we expect to have previously detected these planets in \tess{} data if they exist. The non-detection of these predicted inner planets can be explained if these planets do not transit, which would suggest large mutual inclinations across the \tess{} multi-planet sample. While large mutual inclinations are seen in the \textit{Kepler} compact multi-planet systems with the shortest-period inner planets \citep[$a/R_* < 5$,][]{Dai2018}, this phenomenon is unlikely to impact the \tess{} multi-planet sample, as 95$\%$ of these systems' inner planets have $a/R_* > 5$. Therefore, the fact that these inner planets have not been previously detected by \tess{} suggests that the PCM is not accurately predicting new planets in these systems. 

\begin{figure*}
    \includegraphics[width=\textwidth]{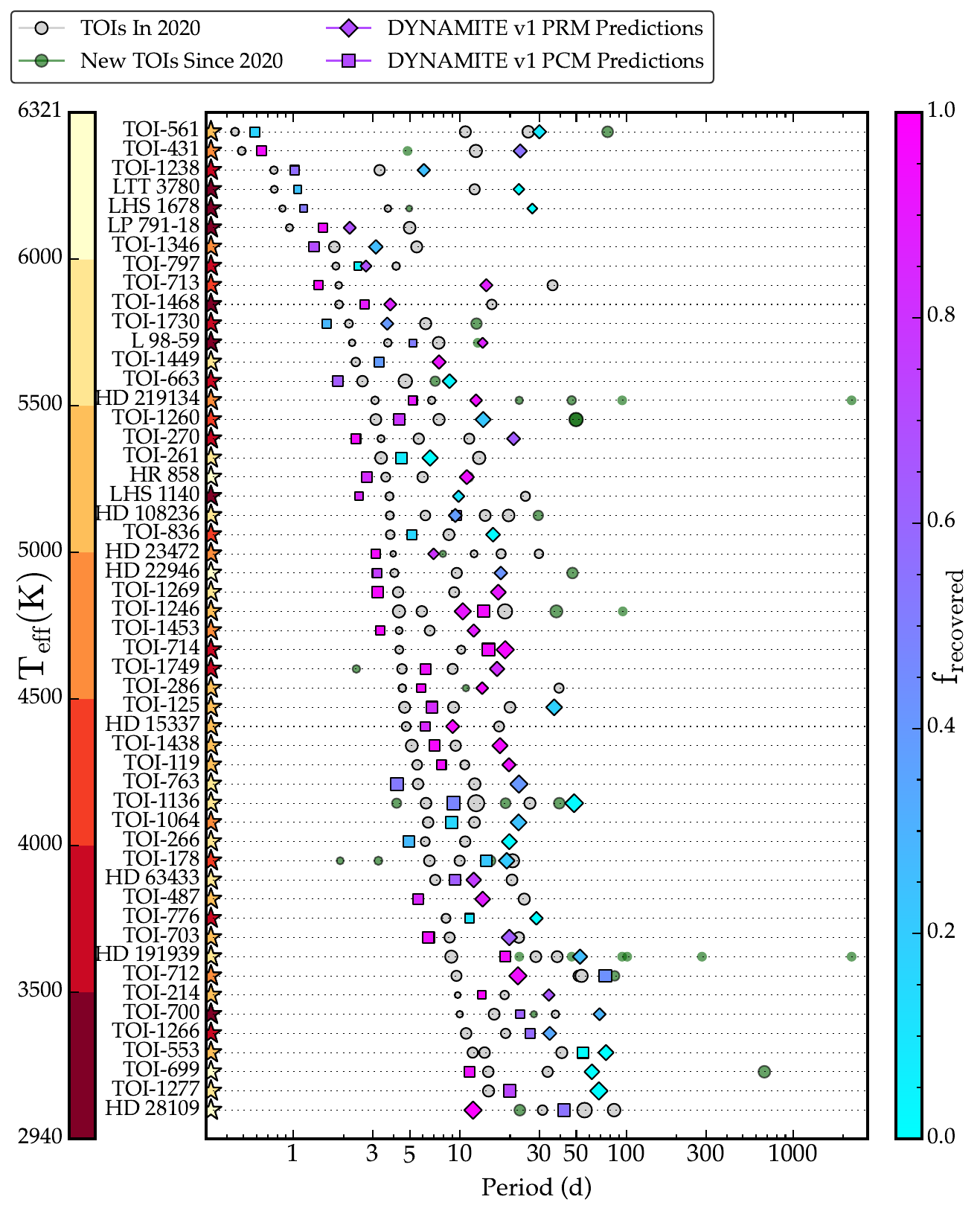}
    \caption{52 \tess{} systems in the 2020 sample with multiple planets and/or confirmed planets (grey circles), with the peak of the \tnt{} v1 orbital period posterior distributions reported in D20 under the period ratio model (PRM, diamonds) and clustered period model (PCM, squares) shaded according to their recovery fraction. The PCM predictions lie at systematically shorter periods, and have higher recovery fractions than the PRM predictions. Planets and planet candidates discovered since 2020 are shown by green circles (outlined in black if new planets have measured radii).}
    \label{fig:injrec_results2020}
\end{figure*}

\begin{figure*}
    \includegraphics[width=\textwidth]{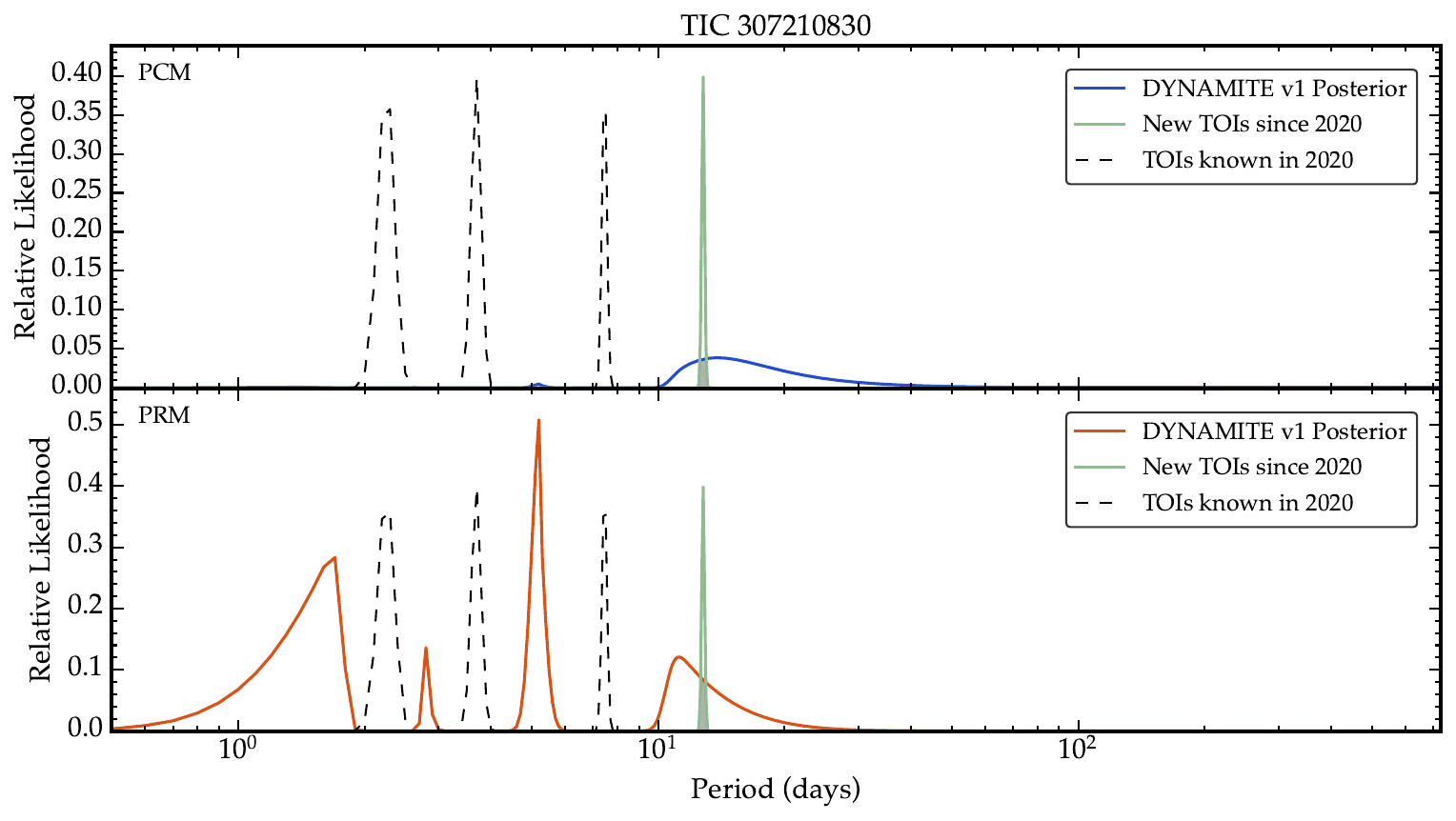}
    \caption{\tnt{} v1 posterior period probability distributions (\citealt{dietrich+22}) for the clustered (PCM, top, blue) and period ratio (PRM, bottom, orange) models for TOI-175 (TIC 307210830, L 98-59). The \tnt{} posteriors are multiplied by a factor of 10 for improved visual clarity. The orbital period distributions (assuming Gaussian errors) of TOIs known in 2020 are shown in black dashed lines, and those that have since been discovered are shown in green. The overlap (shaded in grey) between the newly discovered planet's period and the PRM posterior is 1.5 times larger than that with the PCM posterior. However, in this case, the period of L 98-59 e is closer to that of the highest peak in the PCM posterior; the highest probability period for a new planet under the PRM was at $\sim5$ d.}
    \label{fig:overlap}
\end{figure*}

\begin{figure*}
    \includegraphics[width=22cm,angle=90]{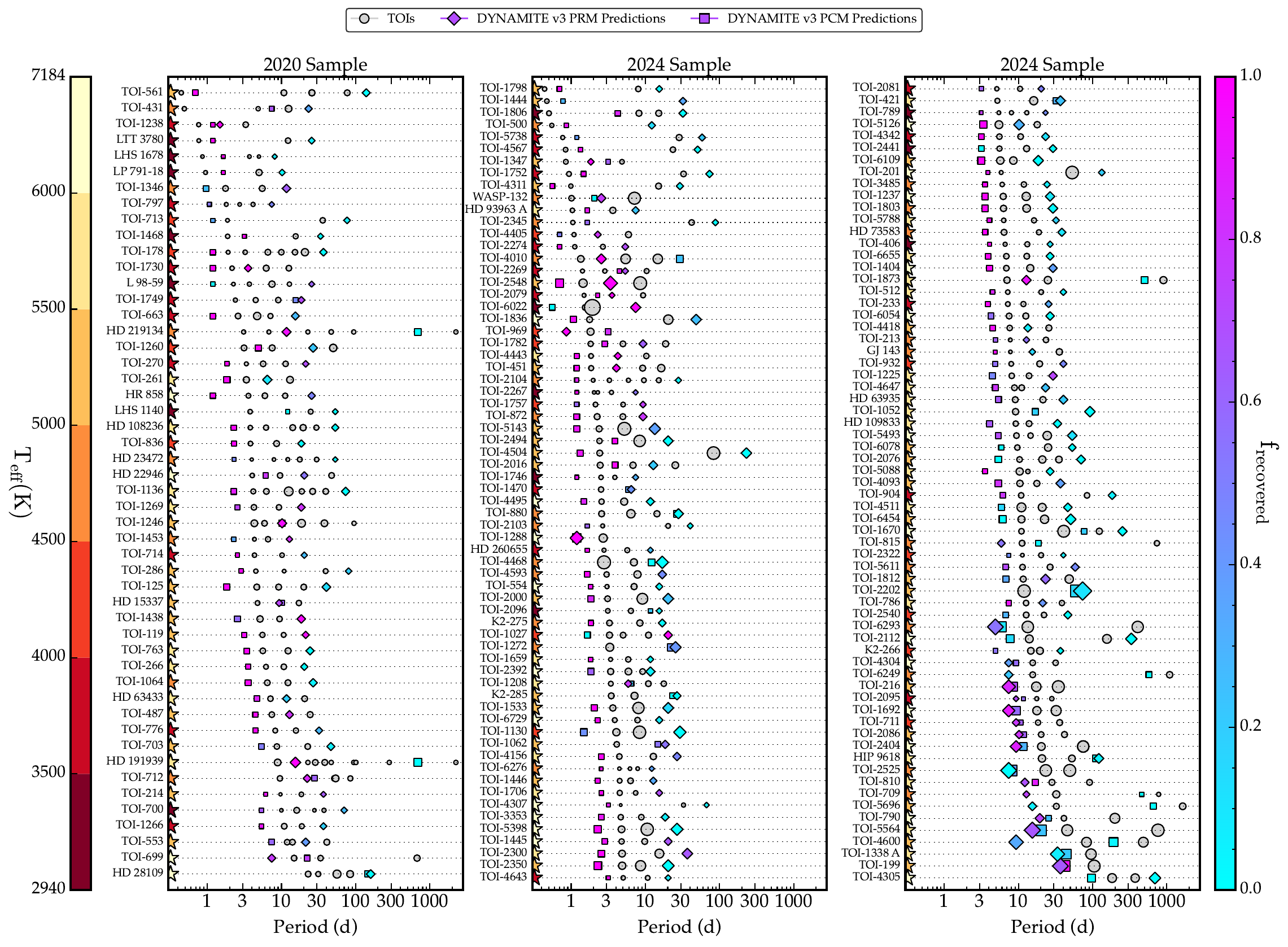}
    \caption{183 \tess{} systems with multiple planet candidates and/or confirmed planets (grey circles showing transiting planets), with the peak of the \tnt{} v3 orbital period posterior distributions under the period ratio model (PRM, diamonds) and clustered period model (PCM, squares) shaded according to their recovery fraction. The PCM predictions lie at systematically shorter periods, and have higher recovery fractions than the PRM predictions.}
    \label{fig:injrec_results2023}
\end{figure*}

\section{Discussion} \label{sec:discussion}

\subsection{How accurately did the PRM and PCM predict additional planets?}
We found that of the \Nnewtrpl{} transiting planets discovered since 2020 in the 2020 sample, the orbital period distributions of 8 have $>10\%$ overlaps with either of the PCM and PRM period posteriors reported in D20. Furthermore, 13 of the newly discovered planets have $>10\%$ overlaps with the PCM and/or PRM predictions using \tnt{} v3 \citep{dietrich+22}.

Using \tnt{} v1, which assumed circular orbits, the posteriors under the PRM are more consistent with the newly discovered planets and planet candidates than those under the PCM. Using the \tnt{} v3 PRM and PCM posteriors, which include eccentricity and inclination, we find that the PRM and PCM performed comparably well. We also found that the periods of the newly discovered planets are closer to the primary peak in the PRM period posteriors than the primary peak in the PCM posteriors (see Fig. \ref{fig:overlap}).  

In cases where multiple additional planets have been discovered in a system, we calculated each overlap separately, as each newly detected planet is an additional check of the models' performances. This result is modulated by the fact that that the planets predicted under the period ratio model were typically at longer orbital periods ($P_{median, PCM} = 3.55$ days, $P_{median, PRM} =$ 20.3 days, see Fig. \ref{fig:injrec_results2020}) and thus both less detectable in a given light curve and less likely to transit. The longer periods associated with the PRM predictions are due to the power law distribution used to describe the innermost planet in each system, which strongly peaks at 10 days \citep[see Fig. 8 in][]{Mulders2015a}. On the other hand, the PCM implements an almost uniform distribution between 3 and 30 days \citep[see Fig. 5 in][]{he+2019}. 

However, the improvement in performance between the \tnt{} v1 implementations of the PRM and PCM are not statistically significant; both models have limited predictive power. The predictions from both models are more accurate when \tnt{} v3 is used, but only a minority of systems are accurately described by PRM or PCM using either version of \tnt{}. This results showcases the limitations in these empirical models; they are not representative of all observed multi-planet systems.

\begin{figure}
    \includegraphics[width=\columnwidth]{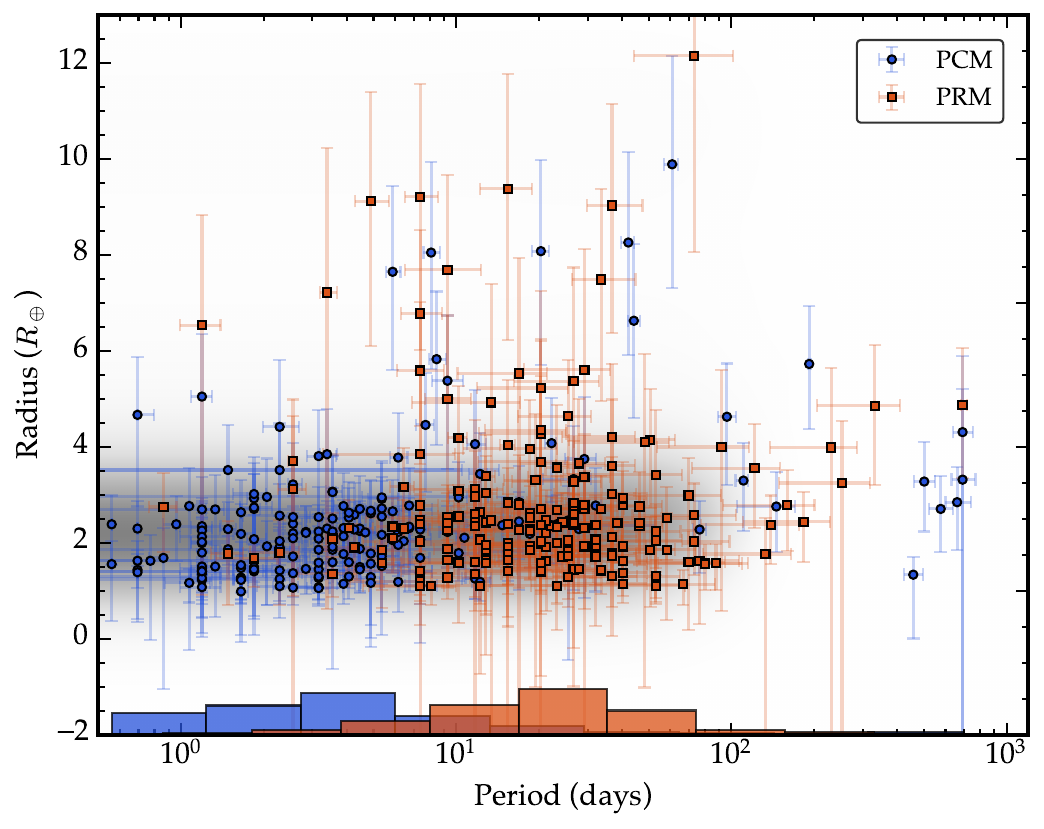}
    \caption{Predicted planet orbital period and radius peaks of the \tnt{} v3 posterior  distributions (PCM: blue circles, PRM: orange squares) for the 2024 sample of 183 \tess{} multi-planet systems. A Gaussian kernel density estimate of the period and radius distributions of the known planets and planet candidates in these systems is shown in grey. The distribution of the orbital periods of the predicted additional planet in each system under the two models are shown as histograms at the bottom of the plot. The period ratio model predicts additional planets at longer orbital periods than the clustered period model.}
    \label{fig:rp}
\end{figure}

\subsection{How does the \tess{} observing strategy impact detectability?}
Our sensitivity to detected additional transiting planets in multi-planet systems was greatly impacted by the observing pattern of the \tess{} mission. During its prime mission, \tess{} observed first the southern and then the northern ecliptic hemispheres, repeating this pattern in the First Extended mission (which also included some pointings along the ecliptic plane). In its Second Extended mission, which began in September 2022, \tess{} has generally observed the southern ecliptic hemisphere. This observing strategy has led to many targets being observed for a few sectors every other year; the median duty cycle of our targets is only 13$\%$, and each light curve has two gaps (corresponding to cycles observing the other hemisphere) which are at least 14 (and up to 40) sectors long. These large gaps substantially inhibit our sensitivity to detecting additional planets. Even TOIs with orbital periods as short as half a sector length (roughly 13 days) may be observed in transit only once per sector due to the timing of transits relative to data gaps. These transit events lead to poorly constrained orbital periods, and require intensive photometric follow-up to characterize \citep[e.g.][]{tuson+23, ulmer-moll+22}. Additionally, transiting planets with periods longer than a sector length ($\sim30$ days) will not transit multiple times each time \tess{} is observing the system, resulting in mono- and duo-transits across the full light curve. The planet detection pipeline used in this work searches for periodic transit-like signals; it is not sensitive to mono-transits. 

\begin{figure}
    \includegraphics[width=\columnwidth]{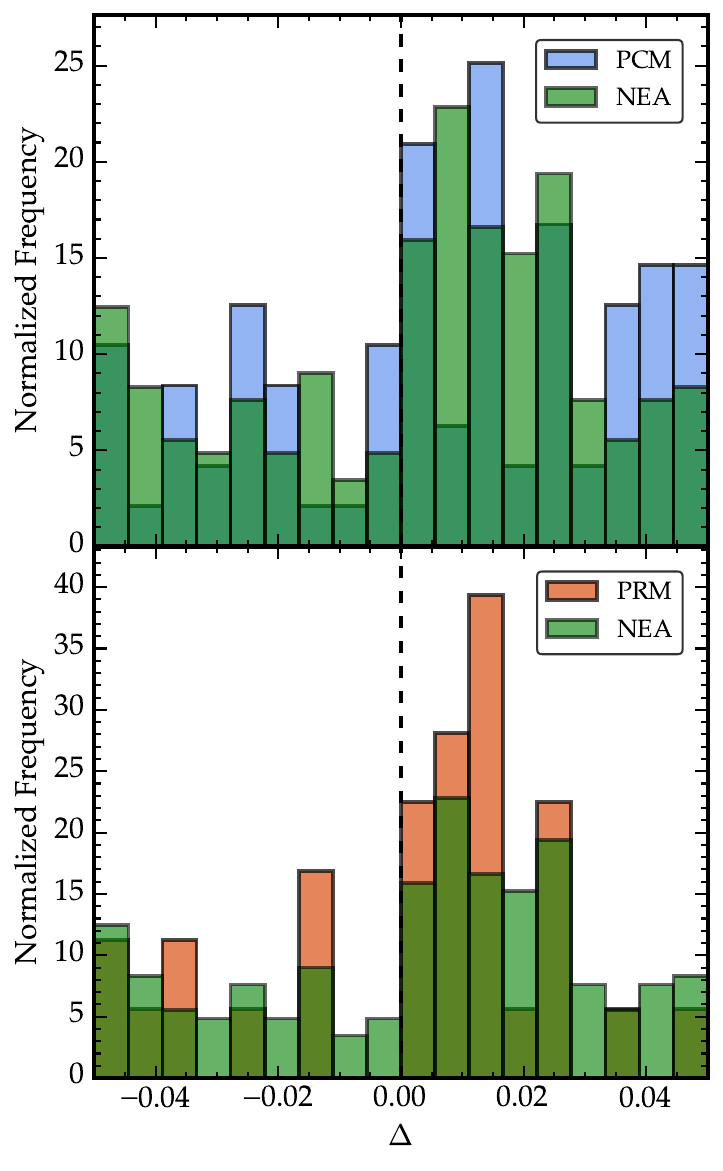}
    \caption{Histograms showing the normalized frequency of adjacent planet pairs' distance from mean-motion resonance \citep[$\Delta$, see Equation \ref{eq:2},][]{lissauer+2011}. The \tnt{} v3 predictions for additional planets in the 52 \tess{} multi-planet systems are included for the period ratio model (blue, upper panel) and the clustered period model (orange, lower panel). The sample of confirmed multi-planet systems in the NASA Exoplanet Archive$^1$ are shown for reference (green, both panels). Both the PCM and PRM models over-predict the excess of planet pairs wide of resonance, and the period ratio model under-predicts planet pairs interior to resonance.}
    \label{fig:twopanel_deltas}
\end{figure}

\subsection{How do \tess{} multi-planet systems compare to the confirmed planet population?} \label{sec:cps}

The models used to predict additional planets in these \tess{} multi-planet systems are based on the observed \textit{Kepler} sample of multi-planet systems. As such, it is important to compare the two stellar samples. We used the one-sided Kolmogorov-Smirnov test \citep{smirnov1939estimation} to evaluate the likelihood of our sample stellar properties being drawn from the Gaia-\textit{Kepler} Stellar Properties Catalog \citep[GKSPC,][]{berger+20}, the sample of confirmed planet and planet candidate \textit{Kepler} Objects of Interest (KOIs\footnote{ExoFOP-KOIs, \url{https://exofop.ipac.caltech.edu/tess/view_koi.php}, accessed 30/04/2024}), and the sample of \textit{Kepler} multi-planet systems$^{4}$. The GKSPC homogenously derived stellar properties for 186,301 \textit{Kepler} stars, using isochrones, broadband photometry, \textit{Gaia} DR2 parallaxes, and spectroscopic metallicities. When considering stellar mass, radius, effective temperature, surface gravity, and metallicity ([Fe/H]), we found that we can reject the null hypothesis (that our stars were drawn from the GKSPC distribution) in all cases with p-value $< 10^{-10}$. We can also reject the null hypotheses when comparing the \tess{} multi-planet system stellar sample with the KOI and multi-planet KOI samples with p-values $<10^{-13}$. However, when comparing the the KOI samples with the full GKSPC, the null hypothesis is also rejected with p-value $<10^{-19}$. This highlights the impact of detection biases; planets orbiting bright, hot stars are most easily detected and followed up. Nevertheless, we find that the stellar properties in our sample are not representative of the wider \textit{Kepler} planet host star sample used to develop the predictive models. This result highlights the importance of acknowledging the influence of stellar properties on trends in exoplanetary systems \citep[e.g.][]{berger+20, fulton+18, fischer+05}.

We also compared the period distribution of our sample of planets (including additional predicted planets) to the period distribution of the confirmed planet population\footnote{NASA Exoplanet Archive, accessed 30/04/2024} in Fig. \ref{fig:twopanel_deltas}. All of the distributions reproduce the known excess of planet pairs wide of resonance \citep{lissauer+2011, lithwick+12a, fabrycky+2014}, as measured using 
\begin{equation} \label{eq:2}
    \Delta = \frac{P_2}{P_1} \frac{j - 1}{j} - 1
\end{equation}
where $P_2 > P_2$, and for $j:j-1$ resonances \citep{lithwick+12b}. We note that this $\Delta$ is distinct from that discussed in Section \ref{sec:packing}, which refers to the physical spacing between planets. However, compared to the distribution for confirmed planets, the period ratio distribution under the period ratio model over-predicted the frequency of pairs wide of resonance, and under-predicted pairs interior to first-order mean-motion resonances. These discrepancies imply that the evolution of observed planets under this model would either involve enhanced near-resonant pair formation \citep[e.g.][]{choksi+chiang20}, or reduced disruption of resonant pairs \citep[e.g.][]{izidoro+21}. 

\subsection{What is the discovery space for additional planets?}
\begin{figure*}
    \includegraphics[width=\textwidth]{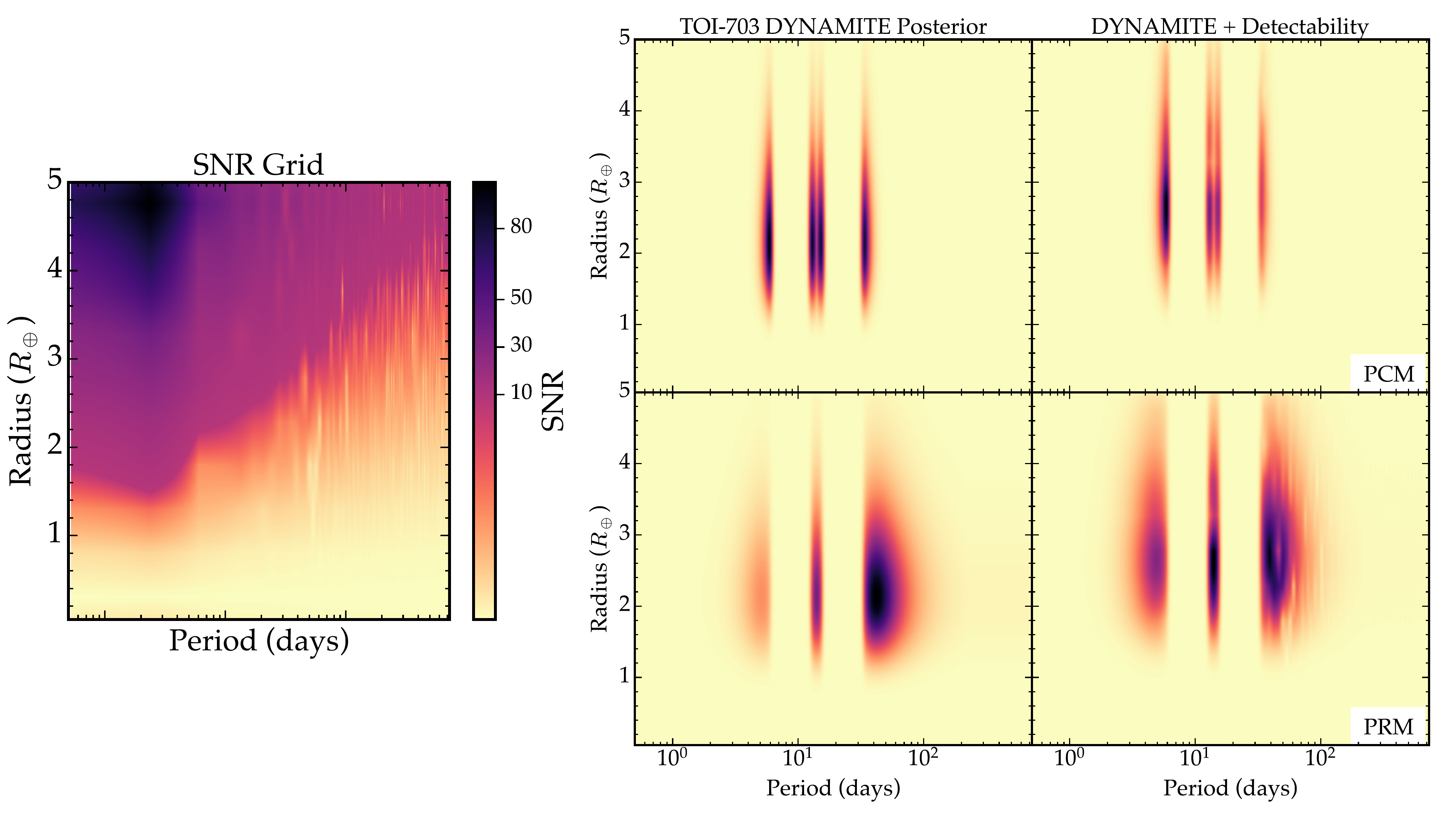}
    \caption{\textit{Left:} Signal-to-Noise Ratio (SNR) of simulated transits in the \tess{} light curve of TOI-703 with corresponding color bar, where the transitiong point in the color map is at 7.3$\sigma$. \textit{Middle:} Mesh of radius and period posteriors generated by \tnt{} v3 for TOI-703. \textit{Right:} Detectability modulated by the \tnt{} v3 posterior for TOI-703, showing the location in ($P, R_p$) space where an additional planet in this system is most likely to be detected under the period ratio model}
    \label{fig:conv}
\end{figure*}

The majority ($66\%$) of the \tess{} multi-planet systems have been or will be re-observed by \tess{} in cycles 6 and 7, as well as other ground-based facilities collecting both photometry and spectroscopy. We can make use of the posterior probability distributions returned by \tnt{} v3 and the sensitivity distributions calculated in this work to evaluate regions of parameter space where additional planets are both most likely to exist and be detected by \tess{}. In order to do so, we generated grids of detectability in period-radius space, parameterized by the SNR of the transits of a putative planet with a given ($R_p, P$) in the light curve of a given star using Equation \ref{eq_snr}. We also computed a mesh of the radius and period posteriors returned by \tnt{} v3 using the dot product. We then down-sampled the posterior grid in order to match the sampling of the SNR grid, and convolved the posterior and detectability grids together. We include a representative example in Fig. \ref{fig:conv}, and make the rest of the plots publicly available on Zenodo\footnote{\url{https://zenodo.org/records/14183270}, \citealt{zenodo}}. The detectability grid modulates the \tnt{} posteriors, demonstrating that additional planets at shorter periods and with larger radii will be most detectable in additional \tess{} observations. These results may be of use in planning follow-up observations of these systems to maximise the return of limited telescope resources. For example, TOI-1339 (HD 191939) is a bright star (V magnitude $\sim9$) that hosts at least 3 transiting and 3 non-transiting planets. The updated \tnt{} v3 orbital period posterior distribution peak at $>50$ days; this region of parameter space not yet been searched well.

\subsection{Are \tess{} multi-planet systems dynamically packed?} \label{sec:packing}
The 2024 sample of 183 \tess{} multi-planet systems host a total of 453 planets and candidate planets. We evaluated the degree to which the systems in this sample are dynamically packed. The Packed Planetary System hypothesis suggests that planet formation is highly efficient, and that as a result, each system will have as many planets as dynamically feasible \citep{barnes+raymond04}. Previous works have shown that the multi-planet systems discovered by \textit{Kepler} are mostly dynamically packed. These works used both N-body simulations \citep{fang+13, pu&wu15} and analytical approaches \citep{humphrey+2022} to evaluate the dynamical stability, and thus packedness, of the \textit{Kepler} multi-planet systems. \citet{fang+13}found that $>31\%, >35\%$ and $>45\%$ of 2-, 3-, 4-planet Kepler systems are dynamically packed on $\sim$Gyr timescales, which is consistent with the $\sim50\%$ of systems found to be packed by \citet{pu&wu15}. A complementary analysis by \citet{humphrey+2022} found that $\sim70\%$ of \textit{Kepler} planet pairs (as opposed to entire systems) were unpacked, or in other words, could host additional planets in existing gaps without becoming dynamically unstable. We sought to analyze the unpacking fraction of the \tess{} multi-planet systems in order to understand not only their stability to additional planets predicted by \tnt{} but also their packedness as currently observed.

As in previous works \citep[see][]{humphrey+2022}, we calculated the minimum and maximum semi-major axes possible for an additional planet in a given system, assuming a minimum spacing (notated as $\Delta_{crit}$) between two adjacent planets. We note that $\Delta_{crit}$ is distinct from the $\Delta$ discussed in Section \ref{sec:cps}. $\Delta$ refers to a planet pair's unit-less distance to mean-motion resonance. $\Delta_{crit}$ refers to the minimum spacing allowable between two adjacent planets if the pair is to remain dynamically stable on $\sim10^8$ yr timescales, and is defined in units of mutual Hill radius \citep{fang+13}. If the difference between the maximum and minimum semi-major axes permissible for an additional planet were $\leq 0$, we considered the known planet pair as packed, and therefore unable to host an additional planet.

For each \tess{} multi-planet system in the 2024 sample, we estimated the mass of the predicted additional planets using the \texttt{spright} \citep[][developed for planets $<5 R_\oplus$]{spright} and \texttt{forecaster} \citep[][]{chen&kipping17} packages. We used \texttt{spright} to predict the masses for the 70 planets/planet candidates in the sample without measured masses. This package uses the planet radius to infer a mass distribution using the mass-radius relationship inferred from either the sample of small transiting confirmed planets orbiting M dwarfs \citep[STPM, ][]{luque+palle22} or orbiting FGK stars \citep[TEPCat\footnote{TEPCat, \url{https://www.astro.keele.ac.uk/jkt/tepcat/}},][]{tepcat}. For the 38 planets or planet candidates in the sample without measured masses and with radii greater than $5R_\oplus$, we used \texttt{forecaster}, as \texttt{spright} is not suitable for planets $> 5 R_\oplus$. \texttt{forecaster} implements the probabilistic mass-radius relationship presented in \citet{chen&kipping17}, which was conditioned using a sample that included both dwarf planets and late-type stars. We also used this method to estimate the masses of the predicted additional planets under both the clustered period and period ratio models, which are listed in Tables \ref{tab:tnt2023pcm} and \ref{tab:tnt2023prm}. We used these predicted masses to evaluate whether the additional predicted planets in each system were dynamically allowed.

We used 4 different values of $\Delta_{crit}$ to facilitate comparisons to previous works: 8 \citep{dyn2020}, 10 (as in \citealt{pu&wu15} for circular orbits), 12.3  (as in \citealt{pu&wu15} for eccentric orbits), and 21.7 \citep{fang+13}. Tables \ref{tab:packingclust} and \ref{tab:packingratio} list the fraction of additional predicted planets for the 2024 sample which would be allowable for a given $\Delta_{crit}$, which we refer to as the ``unpacking fraction'' of these systems. As expected, some of the predicted additional planets would not be allowed in their respective systems with more conservative values of $\Delta_{crit}$.

While it might be expected that all systems would be allowable with $\Delta_{crit} = 8$ as that value was used by \citet{dyn2020}, we note that some systems are not stable even at this low value of $\Delta_{crit}$. This is due to additional planets discovered in several systems since the \tnt{} v3 predictions were made. Additionally \tnt{} used a non-parametric mass-radius relationship \citep{ning+18} to predict planet masses, which returned systematically lower planet masses for given radii than \citet{spright} and \citet{chen&kipping17}. Overall, the majority of both the PCM and PRM predicted additional planets are allowable for $\Delta_{crit} \leq 12.3$.

We also considered the dynamical packing of the 2024 sample of \tess{} multi-planet systems, excluding predicted additional planets. For each pair of adjacent known planets in a system, we calculated whether a $1 M_\oplus$ planet would be dynamically allowable in the gap, for a range of critical separations ($\Delta_{crit}$). We sum this binary variable (i.e., 1 for a dynamically allowed planet, 0 for a dynamically disallowed planet) across all adjacent planet pairs in a system, and divide by the number of planet pairs to return a measure of how allowable an additional Earth-mass planet would be in the system, or the ``unpacking fraction''. Table \ref{tab:packing} summarizes these results; we find that for $\Delta_{crit} = 8$, the majority (73\%) of planet pairs could host an additional intermediate Earth-mass planet, but that this fraction falls to just $21\%$ for the most conservative value of $\Delta_{crit}$ (21.7). We also note that the total unpacking fractions are lower than those reported in Tables \ref{tab:packingclust} and \ref{tab:packingratio}. This is because we require the additional Earth-mass planet to lie between two know planets, whereas the majority of predicted additional planets under each model lie interior to the innermost known planet or exterior to the outermost known planet in the system (see Section \ref{sec:dyn23sample}); these results are not directly comparable. 

When considering the sub-sample of \tess{} multi-planet systems hosting confirmed planets with P $<$ 200 d and 1.5 - 30 $R_\oplus$ and assuming $\Delta_{crit} = 21.7$ (as in \citealt{fang+13}), we find that the observed packing fraction of the observed planets is high. Namely, 100$\%$, 100$\%$, and 75$\%$ of 2-, 3-, and $>$3-planet systems ($N_{2pl} = 52 , N_{3pl} = 11, N_{4pl} = 1$) are packed. Furthermore, 92$\%$ of all systems in this sample are packed. These results are consistent, though substantially higher than those reported in \citet{fang+13}, possibly due to the small number of high-multiplicity systems in our sample. Future analyses which draw $\Delta_{crit}$ from more complex distributions \citep[as suggested by][]{Dietrich2024} rather than a step function may bring the results of the \textit{Kepler} and \tess{} samples into even closer agreement. 

If we apply the constraints used by \citet{pu&wu15} ($N_{pl} > 4$, 183 \tess{} systems), we find that $\sim72\%$, and $83\%$ of systems are packed assuming $\Delta_{crit}$ values of 10 and 12.3, respectively. More recent work by \citet{obertas+23} found that by relaxing assumptions of circular orbits and strictly central placing of additional planets, an even larger fraction ($60-90\%$) of \textit{Kepler} multi-planet systems are strongly packed. The packing fraction that we found for the sub-sample of confirmed \tess{} planets is also high, possibly implying a highly efficient formation history followed by dynamical sculpting to produce the observed lower-multiplicity sample \citep[e.g.][]{ghosh+chatterjee24}.

\subsection{Are \tess{} multi-planet systems truncated?}
We also studied the observed outer edges of these multi-planet systems. \citet{millholland2022} found that $\gtrsim35\%$ of \textit{Kepler} multi-planet systems are truncated. This truncation refers to the lack of detected transiting planets with orbital periods $\sim$100 - 300 days, even when accounting for the reduced transit probability at longer orbital periods. This result implies that the outer regions of these multi-planet systems either have a dearth of planets, or contain only small, and specifically un-detectable, planets. However, multi-planet systems often exhibit relative uniformity in planet radii and masses, also known as the ``peas-in-a-pod'' architecture \citep{millholland+2017, weiss+2018}. Therefore, a decrease in the typical planet radius with increasing orbital period may reveal a break-down of this architecture, which has already been noted in wide multi-planet systems (with at least one planet $>100$ d, \citealt{brewer+23}).

There are several possible mechanisms that could drive this breakdown. Firstly, small planets may be more susceptible to migration traps, preventing them from migrating inwards, unlike larger planets formed in the system \citep{Zawadzki2022}. Secondly, \citet{adams+20} find that for systems with $\gtrsim 40 M_\oplus$ in total planetary mass, mass uniformity is no longer the most energetically favorable architecture for semi-major axes of $\sim0.5-1 AU$. Thirdly, giant outer planets with orbital periods $>300$ days may destabilize planets in this period range \citep[e.g.][]{denham+19}, leading to a dearth of planets in these intermediate regions of multi-planet systems, although this interpretation is unlikely to contribute substantially to the observed truncation \citep{sobski23}. We sought to investigate whether the \tess{} multi-planet systems also exhibit signs of truncation.

We performed a heuristic investigation into the outer edges of the 34 TOIs with 3 or more planets or planet candidates, following the methodology of \citet{millholland2022}. Firstly, we defined a putative external transiting planet in each system, based on the period spacing and radius uniformity seen in peas-in-a-pod architectures. A caveat of this approach is that, as demonstrated in Section \ref{sec:results}, we find that this model of multi-planet systems is not fully descriptive of the observed population. We assigned a radius to the putative planet equal to the radius of the outermost transiting planet, and a period set by the linear spacing of the two outermost transiting planets in the system. The hypothesized external planets range in orbital period from $\sim5$ to 370 days, and in radius from 0.9 to 6.5 $R_\oplus$. In comparison, the hypothesized external planets for the Kepler multi-planet sample ranged in orbital period from $\sim 7$ to 750 days; the Kepler sample probes a longer period population of planets, although we attribute this to the longer observational baselines for Kepler targets.

\begin{figure*}
    \includegraphics[width=\textwidth]{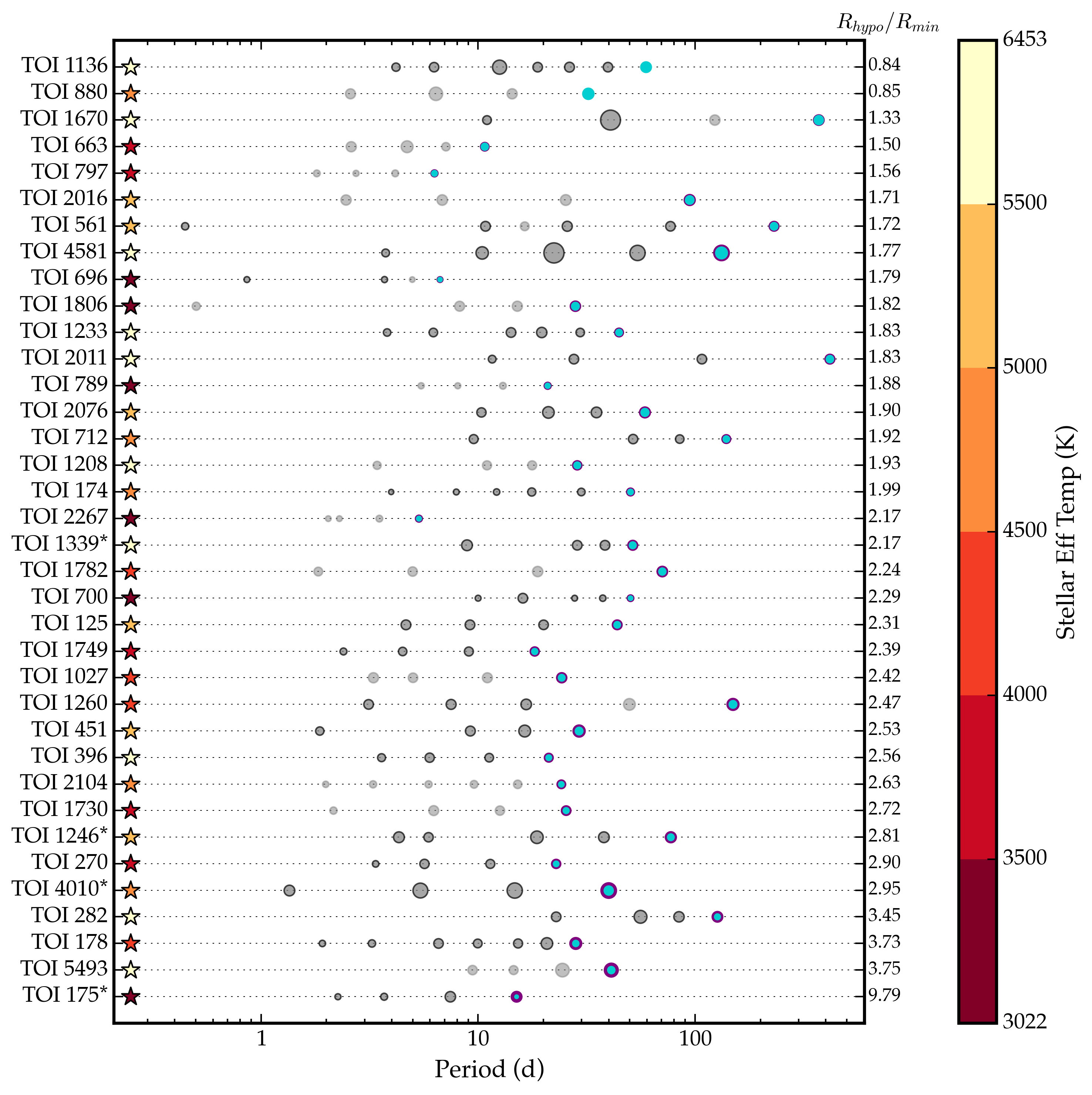}
    \caption{Transiting planets (grey, outlined in black) and planet candidates (grey, no outlines) in the 36 \tess{} systems with 3 or more planets or planet candidates. Systems with additional non-transiting confirmed planets are indicated with a single asterisk. All points are scaled according to planet radius. A putative outer planet (purple) is included for each system, with a hypothetical radius equal to that of the outermost transiting planet in the system. The minimum detectable radius at the hypothetical planet's orbital period is shown in blue. On the right of the plot, the ratio between the hypothetical and minimum detectable radii for the putative planets is shown. On the left of the plot, the effective temperature of the stellar host is shown in the colour of a star-shaped marker. If the purple putative planet is not visible (as is the case for TOIs 1136 and 880), the putative planet is smaller than the minimum detectable radius, and thus would not be detectable in the current \tess{} light curve.}
    \label{fig:edgeofmultis}
\end{figure*}

We then calculated the minimum radius of a putative transiting planet with this orbital period that would be detectable to $>7.3\sigma$ in each system using the full \tess{} data set, using Equation \ref{eq_snr}. Next, we compared the calculated minimum detectable radius with the radius of the putative outer planet. We find that for all but two of the 34 systems in the sample, the ratio of the hypothetical radius and the minimum detectable radius is greater than one. This result implies that the putative transiting outer planets should be detectable in the current \tess{} light curve of the system (see Fig. \ref{fig:edgeofmultis}).

However, there are no detected transiting planets or planet candidates at the periods of the putative outer planets. This indicates that planets at these orbital periods, if they do exist, are either too small to detect using current \tess{} data, or inclined such that they do not transit. We used a combination of literature inclinations for confirmed planets and measured inclinations (using \texttt{TATER}, Mayo et al., in prep) for planet candidates to calculate the mutual inclination of each system. We found that the mutual inclinations of all but one (TOI-2104) of these systems are $\lesssim3^\circ$. This result is consistent with the distribution of mutual inclinations in the \textit{Kepler} multi-planet systems \citep{lissauer+2011, fabrycky+2014}, which is well-described by a Rayleigh distribution with scale parameter $\sigma_i \sim 1-2^\circ$. Additionally, with the exception of Mercury, the orbits of the Solar System planets lie within a few degrees of the ecliptic plane\footnote{NASA Solar System Fact Sheet, \url{https://nssdc.gsfc.nasa.gov/planetary/factsheet/}}. Mutual inclination is a probe of the dynamical temperature of a system; nearly co-planar systems are ``cooler'' than those where planet eccentricities and inclinations have been excited by the Kozai-Lidov mechanism \citep[e.g.][]{kozai62, lidov62, fabrycky+tremaine07} or tidally-driven migration \citep[e.g.][]{jackson+08, wu+24}. For 11 of the 34 high-multiplicity \tess{} systems, the minimum transiting inclination of the putative outer planet implies that if an additional, evenly-spaced outer planet were to exist and not transit, these systems would be dynamically hotter than currently understood. 

The \tess{} multi-planet systems appear to be truncated relative to detection sensitivity limits when considering transiting planets, similarly to the \textit{Kepler} systems. However, we cannot test the hypothesis of the dearth of planets in the 100-300 d orbital period range described in \citet{millholland2022} due to the heterogeneous and shorter photometric baselines of \tess{} systems compared to the \textit{Kepler} sample. Only 5/34 ($15\%$) of the systems in the \tess{} high-multiplicity sample had putative outer planets with orbital periods longer than 100 days, as opposed to 20/64 Kepler systems ($31\%$). Additionally, a two-sided Kolmogorov-Smirnov \citep{smirnov1939estimation} test of the host star effective temperature samples revealed that the \tess{} (3022 - 6453 K) and Kepler (3740 - 6360 K) samples are not consistent with the same underlying distribution. As such, the truncation of multi-transiting planet systems may be independent of stellar population.

\section{Conclusions} \label{sec:conclusion}
We used an additional 4 years (54 sectors) of \tess{} observations to search for predicted additional planets in each of 52 multi-planet systems, under two predictive models based on results from the \textit{Kepler} mission. We found that a period ratio model more accurately predicted additional planets in these multi-planet systems. However, neither the period ratio model \citep{mulders+2018} nor a clustered period model \citep{he+2019} accurately predicted all of the additional planets found in these multi-planet systems. This result highlights the need for additional data to continue to improve models of multi-planet systems, as well as the need for prescriptions for eccentricity and inclination, as in \citet{dietrich+22}.

We use an updated set of predictive models from \tnt{}, \citep{dietrich+22} to iteratively predict additional planets in the 183 current \tess{} systems with multiple planets or planet candidates. We modulate these posterior probabilities with these putative planets' detectability and provide the most detectable planet location in orbital period and radius space. The majority of these systems will be re-observed by \tess{} in Cycle 7, and the remainder will likely be re-observed in the third extended mission. These additional observations will enable further testing and refinement of the models discussed here, and continue to expand our understanding of these multi-planet systems.

We also found that the high-multiplicity ($N_{pl} \geq 3$) transiting systems observed by \tess{} appear to be truncated and highly dynamically packed. While the nature of the \tess{} mission makes homogeneous analyses analogous to those done with \textit{Kepler} data challenging, continued and sustained follow-up efforts of these rich systems is warranted to understand how these planets formed and came to be as they are observed today.

\begin{acknowledgments}
We thank Guy Nir and Massimo Pascale for useful conversations in the preparation of this work. E. Turtelboom acknowledges support from a David \& Lucile Packard Foundation grant. C. Harada acknowledges support from the National Science Foundation Graduate Research Fellowship Program under Grant No. DGE 2146752. C.D.Dressing gratefully acknowledges support from the NASA Exoplanets Research Program (XRP) via grant 80NSSC20K0250, the Hellman Faculty Fellows Fund, the Sloan Foundation, and the David \& Lucile Packard Foundation.

This research has made use of the Exoplanet Follow-up Observation Program (ExoFOP; DOI: \dataset[10.26134/ExoFOP5]{http://dx.doi.org/10.26134/ExoFOP5}, \citealt{exofop5}) website, which is operated by the California Institute of Technology, under contract with the National Aeronautics and Space Administration under the Exoplanet Exploration Program. This paper made use of data collected by the \tess{} mission and are publicly available from the Mikulski Archive for Space Telescopes (MAST) operated by the Space Telescope Science Institute (DOI: \dataset[10.17909/fwdt-2x66]{http://dx.doi.org/10.17909/fwdt-2x66}, \citealt{tic}). Funding for the \tess{} mission is provided by NASA’s Science Mission Directorate. We acknowledge the use of public \tess{} data from pipelines at the \tess{} Science Office and at the \tess{} Science Processing Operations Center. Resources supporting this work were provided by the NASA High-End Computing (HEC) Program through the NASA Advanced Supercomputing (NAS) Division at Ames Research Center for the production of the SPOC data products. This research has made use of NASA’s Astrophysics Data System, and of the NASA Exoplanet Archive (DOI: \dataset[10.26133/NEA1]{http://dx.doi.org/10.26133/NEA1}, \citealt{nea1}). 

\end{acknowledgments}

\facilities{\tess{}}

\software{astropy \citep{astropy:2013, astropy:2018, astropy:2022}, batman \citep{batman}, lightkurve \citep{lk}, wotan \citep{wotan}}

\bibliography{main}
\bibliographystyle{aasjournal}

\appendix

In this Appendix, we include Tables \ref{tab:packingclust}, \ref{tab:packingratio}, \ref{tab:packing}, which include the fraction of the 2024 sample of multi-planet systems for which the additional predicted planets under the clustered period model, the additional predicted planets under the period ratio model, and additional 1 $M_\oplus$ planets would be allowed for a range of critical separations ($\Delta_{crit}$). We also include Tables \ref{tab:summarydynv1} and \ref{tab:summarydynv3}, which detail the \tnt{} v1 and v3 predictions based on \tess{} Sectors 1-26 for the 52 \tess{} multi-planet systems as well as the new planet candidates discovered in these systems since 2020. Finally, we include Tables \ref{tab:tnt2023pcm} and \ref{tab:tnt2023prm}, which detail the up-to-date \tnt{} predictions based on \tess{} sectors 1-76 for the current sample of 183 \tess{} multi-planet systems.

\include{summary_DYNv1}
\include{summary_DYNv3}

\include{dynv3predsPCM}
\include{dynv3predsPRM}

\include{packing_table_clust}
\include{packing_table_ratio}
\include{packing_table_nopred}

\end{document}

%% file: authors.tex
\author[0000-0002-1845-2617]{Emma V. Turtelboom}\affiliation{Department of Astronomy, 501 Campbell Hall, University of California, Berkeley, CA 94720, USA}

\author[0000-0001-6320-7410]{Jamie Dietrich}
\affiliation{School of Earth and Space Exploration, Arizona State University, Tempe, AZ 85287, USA}

\author[0000-0001-8189-0233]{Courtney D. Dressing}\affiliation{Department of Astronomy, 501 Campbell Hall, University of California, Berkeley, CA 94720, USA}

\author[0000-0001-5737-1687]{Caleb K. Harada} \altaffiliation{NSF Graduate Research Fellow} \affiliation{Department of Astronomy, 501 Campbell Hall, University of California, Berkeley, CA 94720, USA}

%% file: summary_DYNv1.tex
\begin{longrotatetable}
\begin{deluxetable*}{LCC|CCCCCC|CCCCCC|CCCC}
\tablecolumns{19}
\tablefontsize{\tiny}
\tablecaption{Summary of the \tnt{} version 1 (v1) predictions \textbf{with the associated 1$\sigma$ errors} under the period ratio and clustered period models for the 52 TESS multi-planet systems known in 2020. For each model, the period and radius at the maximum of the posterior probability distributions are reported. We also include the TOIs discovered in these systems since 2020, with associated period and radius reported on ExoFOP. Two of the systems (TOIs 1730 and 1749) have had multiple new planet candidates discovered since 2020, which are shown in separate rows.}
\label{tab:summarydynv1}
\tablehead{
    \colhead{TIC} &
    \colhead{TOI} &
    \colhead{Name} &
    \multicolumn{6}{c}{Clustered Period Model} & 
    \multicolumn{6}{c}{Period Ratio Model} & 
    \multicolumn{4}{c}{New TOIs Since 2020} \\ [-5ex] 
    \colhead{} &
    \colhead{} &
    \colhead{} &
    \colhead{Period} &
    \colhead{+ Error} &
    \colhead{- Error} &
    \colhead{Radius} &
    \colhead{+ Error} &
    \colhead{- Error} &
    \colhead{Period} &
    \colhead{+ Error} &
    \colhead{- Error} &
    \colhead{Radius} &
    \colhead{+ Error} &
    \colhead{- Error} &
    \colhead{TOI} &
    \colhead{Period\tablenotemark{a}} &
    \colhead{Radius} & 
    \colhead{Error} \\ [-5ex] 
    \colhead{} & \colhead{} & \colhead{} &
    \colhead{(days)} &
    \colhead{(days)} &
    \colhead{(days)} &
    \colhead{($\mathrm{R_\oplus}$)} & 
    \colhead{($\mathrm{R_\oplus}$)} & 
    \colhead{($\mathrm{R_\oplus}$)} & 
    \colhead{(days)} &
    \colhead{(days)} &
    \colhead{(days)} &
    \colhead{($\mathrm{R_\oplus}$)} & 
    \colhead{($\mathrm{R_\oplus}$)} & 
    \colhead{($\mathrm{R_\oplus}$)} & 
    \colhead{} & \colhead{(days)} &
    \colhead{($\mathrm{R_\oplus}$)} &
    \colhead{($\mathrm{R_\oplus}$)} \\ [-5ex] 
    }
\startdata
278683844&119&$\mathrm{TOI-119}$&7.73&0.41&0.43&1.91&0.66&0.5&19.7&7.96&3.62&1.91&0.66&0.49&&&&\\
52368076&125&$\mathrm{TOI-125}$&6.81&0.0&0.56&2.74&0.91&0.7&36.8&14.9&6.77&2.75&0.9&0.7&&&&\\
425997655&174&$\mathrm{HD 23472}$&3.12&0.0&0.22&1.53&1.0&0.6&6.95&1.14&0.98&1.29&0.45&0.33&174.05&7.91&0.97&0.07\\
307210830&175&$\mathrm{L 98-59}$&5.22&0.03&0.0&1.15&0.4&0.3&13.7&5.55&2.52&1.15&0.4&0.3&&&&\\
251848941&178&$\mathrm{TOI-178}$&14.4&4.78&0.0&2.67&0.89&0.68&19.1&7.71&3.51&2.67&0.88&0.68&178.04&3.24&1.51&1.69\\
167415965&214&$\mathrm{TOI-214}$&13.5&0.77&1.23&1.46&0.87&0.55&34.2&13.8&6.28&1.46&0.88&0.54&214.02&9.7&0.81&0.07\\
92226327&256&$\mathrm{LHS 1140}$&2.48&0.0&0.11&1.45&0.5&0.37&9.81&4.03&3.05&1.45&0.5&0.37&&&&\\
63898957&261&$\mathrm{TOI-261}$&4.46&0.3&0.0&2.62&0.87&0.67&6.62&1.51&1.23&2.62&0.87&0.67&&&&\\
\enddata
\tablenotetext{a}{Typical errors on the period of TOIs reported on ExoFOP are $<0.001$.}
\end{deluxetable*}
\end{longrotatetable}

%% file: summary_DYNv3.tex
\begin{longrotatetable}
\begin{deluxetable*}{LCC|CCCCCC|CCCCCC|CCCC}
\tablecolumns{19}
\tablefontsize{\tiny}
\tablecaption{Summary of the \tnt{} version 3 (v3) predictions \textbf{with the associated 1$\sigma$ errors} under the period ratio and clustered period models for the 52 TESS multi-planet systems known in 2020. For each model, the period and radius at the maximum of the posterior probability distributions are reported. We also include the TOIs discovered in these systems since 2020, with associated period and radius reported on ExoFOP. Two of the systems (TOIs 1730 and 1749) have had multiple new planet candidates discovered since 2020, which are shown in separate rows.}
\label{tab:summarydynv3}
\tablehead{
    \colhead{TIC} &
    \colhead{TOI} &
    \colhead{Name} &
    \multicolumn{6}{c}{Clustered Period Model} & 
    \multicolumn{6}{c}{Period Ratio Model} & 
    \multicolumn{4}{c}{New TOIs Since 2020} \\ [-5ex] 
    \colhead{} &
    \colhead{} &
    \colhead{} &
    \colhead{Period} &
    \colhead{+ Error} &
    \colhead{- Error} &
    \colhead{Radius} &
    \colhead{+ Error} &
    \colhead{- Error} &
    \colhead{Period} &
    \colhead{+ Error} &
    \colhead{- Error} &
    \colhead{Radius} &
    \colhead{+ Error} &
    \colhead{- Error} &
    \colhead{TOI} &
    \colhead{Period\tablenotemark{a}} &
    \colhead{Radius} & 
    \colhead{Error} \\ [-5ex] 
    \colhead{} & \colhead{} & \colhead{} &
    \colhead{(days)} &
    \colhead{(days)} &
    \colhead{(days)} &
    \colhead{($\mathrm{R_\oplus}$)} & 
    \colhead{($\mathrm{R_\oplus}$)} & 
    \colhead{($\mathrm{R_\oplus}$)} & 
    \colhead{(days)} &
    \colhead{(days)} &
    \colhead{(days)} &
    \colhead{($\mathrm{R_\oplus}$)} & 
    \colhead{($\mathrm{R_\oplus}$)} & 
    \colhead{($\mathrm{R_\oplus}$)} & 
    \colhead{} & \colhead{(days)} &
    \colhead{($\mathrm{R_\oplus}$)} &
    \colhead{($\mathrm{R_\oplus}$)} \\ [-5ex] 
    }
\startdata
278683844&119&$\mathrm{TOI-119}$&3.16&0.4&0.5&1.86&0.62&0.48&20.3&9.6&5.7&1.91&0.66&0.49&&&&\\
52368076&125&$\mathrm{TOI-125}$&2.55&0.4&0.5&2.69&0.93&0.69&38.5&14.8&8.6&2.77&0.95&0.71&&&&\\
425997655&174&$\mathrm{HD 23472}$&7.39&1.1&1.2&1.27&0.45&0.32&7.39&0.9&0.9&1.28&0.45&0.32&174.05&7.91&0.97&0.07\\
307210830&175&$\mathrm{L 98-59}$&1.19&11.0&6.0&1.12&0.38&0.28&15.4&6.5&3.7&1.15&0.4&0.29&&&&\\
251848941&178&$\mathrm{TOI-178}$&3.39&0.5&0.7&2.66&0.91&0.68&20.3&8.3&5.1&2.68&0.93&0.68&178.04&3.24&1.51&1.69\\
167415965&214&$\mathrm{TOI-214}$&5.87&0.7&0.9&1.28&0.89&0.41&36.8&14.4&8.4&1.44&0.88&0.52&214.02&9.7&0.81&0.07\\
92226327&256&$\mathrm{LHS 1140}$&2.05&0.1&0.2&1.26&0.43&0.3&9.29&2.0&1.7&1.43&0.5&0.36&&&&\\
63898957&261&$\mathrm{TOI-261}$&5.36&1.55&1.05&2.45&0.78&0.62&6.44&0.7&0.6&2.63&0.91&0.68&&&&\\
\enddata
\tablenotetext{a}{Typical errors on the period of TOIs reported on ExoFOP are $<0.001$.}
\end{deluxetable*}
\end{longrotatetable}

%% file: dynv3predsPCM.tex
\begin{longrotatetable}
\begin{deluxetable*}{LCCCCCCCC}
\tablecolumns{9}
\tablefontsize{\scriptsize}
\tablecaption{\tnt{} v3 predictions for the 2024 dataset with the PCM model}
\label{tab:tnt2023pcm}
\tablehead{
    \colhead{Name} &
    \colhead{Period (d)} &
    \colhead{Radius ($R_\oplus$)} &
    \colhead{Mass ($M_\oplus$)} &
    \colhead{Inclination (deg) } &
    \colhead{Eccentricity} &
    \colhead{Transit Depth (ppm)} &
    \colhead{Transit Probability} &
    \colhead{RV Semi-amplitude (m/s)}
}
\startdata
HD 22946 & $15.4^{+1.8}_{-1.3}$ & $1.96^{+0.65}_{-0.5}$ & $5.04^{+2.88}_{-1.87}$ & $90.5^{+42.6}_{-43.4}$ & $0.036^{+0.029}_{-0.016}$ & $257^{+171}_{-131}$ & $0.523^{+0.093}_{-0.001}$ &$1.22^{+0.26}_{-0.69}$\\
TOI-1873 & $502.1^{+42.7}_{-47.9}$ & $3.28^{+1.04}_{-0.83}$ & $11.37^{+6.2}_{-4.2}$ & $90.0^{+43.1}_{-43.2}$ & $0.037^{+0.03}_{-0.017}$ & $961^{+617}_{-496}$ & $0.502^{+0.007}_{-0.001}$ &$0.91^{+0.25}_{-0.53}$\\
HD 191939 & $692.2^{+52.6}_{-61.7}$ & $4.31^{+9.09}_{-1.58}$ & $17.5^{+87.55}_{-8.99}$ & $88.6^{+1.2}_{-0.8}$ & $0.017^{+0.016}_{-0.008}$ & $1770^{+7450}_{-1300}$ & $0.15^{+0.346}_{-0.001}$ &$1.46^{+7.86}_{-0.79}$\\
TOI-797 & $7.39^{+2.1}_{-1.3}$ & $1.17^{+0.41}_{-0.29}$ & $1.55^{+2.81}_{-0.968}$ & $90.0^{+43.1}_{-42.5}$ & $0.042^{+0.038}_{-0.02}$ & $513^{+361}_{-256}$ & $0.52^{+0.047}_{-0.001}$ &$0.84^{+1.07}_{-0.63}$\\
TOI-2404 & $399.3^{+24.2}_{-30.5}$ & $4.06^{+2.64}_{-1.13}$ & $15.92^{+19.21}_{-6.41}$ & $90.0^{+42.7}_{-43.1}$ & $0.033^{+0.026}_{-0.015}$ & $995^{+1300}_{-561}$ & $0.502^{+0.013}_{-0.001}$ &$1.25^{+1.03}_{-0.76}$\\
TOI-4418 & $12.2^{+2.1}_{-1.6}$ & $2.3^{+0.74}_{-0.57}$ & $6.49^{+3.59}_{-2.35}$ & $90.1^{+42.6}_{-42.3}$ & $0.046^{+0.03}_{-0.02}$ & $661^{+433}_{-338}$ & $0.522^{+0.063}_{-0.001}$ &$1.91^{+0.61}_{-1.12}$\\
TOI-561 & $0.694^{+0.0}_{-0.0}$ & $2.3^{+0.77}_{-0.57}$ & $6.49^{+3.75}_{-2.35}$ & $90.0^{+1.1}_{-1.5}$ & $0.021^{+0.019}_{-0.01}$ & $630^{+422}_{-313}$ & $1.0^{+0.001}_{-0.001}$ &$5.4^{+3.41}_{-2.06}$\\
TOI-554 & $4.46^{+0.8}_{-0.7}$ & $1.41^{+1.33}_{-0.43}$ & $2.94^{+26.19}_{-2.11}$ & $90.0^{+42.9}_{-43.0}$ & $0.045^{+0.029}_{-0.019}$ & $82^{+155}_{-50}$ & $0.561^{+0.162}_{-0.001}$ &$0.99^{+7.4}_{-0.81}$\\
\enddata
\end{deluxetable*}
\end{longrotatetable}

%% file: dynv3predsPRM.tex
\begin{longrotatetable}
\begin{deluxetable*}{LCCCCCCCC}
\tablecolumns{9}
\tablefontsize{\scriptsize}
\tablecaption{\tnt{} v3 predictions for the 2024 dataset with the PRM model}
\label{tab:tnt2023prm}
\tablehead{
    \colhead{Name} &
    \colhead{Period (d)} &
    \colhead{Radius ($R_\oplus$)} &
    \colhead{Mass ($M_\oplus$)} &
    \colhead{Inclination (deg) } &
    \colhead{Eccentricity} &
    \colhead{Transit Depth (ppm)} &
    \colhead{Transit Probability} &
    \colhead{RV Semi-amplitude (m/s)}
}
\startdata
TOI-512 & $40.3^{15.2}_{8.9}$ & $1.62^{0.56}_{0.42}$ & $4.75^{8.49}_{3.07}$ & $90.1^{43.0}_{42.8}$ & $0.071^{0.064}_{0.034}$ & $283^{198}_{150}$ & $0.511^{0.02}_{0.001}$ &$0.94^{1.35}_{0.74}$\\
HD 63935 & $40.3^{16.2}_{9.0}$ & $2.93^{1.03}_{0.75}$ & $9.51^{5.8}_{3.55}$ & $90.0^{3.8}_{4.3}$ & $0.072^{0.063}_{0.035}$ & $783^{554}_{406}$ & $0.535^{0.25}_{0.001}$ &$1.87^{1.51}_{0.85}$\\
TOI-6022 & $7.39^{4.1}_{2.5}$ & $3.85^{14.94}_{1.73}$ & $14.64^{164.57}_{8.94}$ & $90.0^{43.0}_{42.7}$ & $0.075^{0.073}_{0.037}$ & $9110^{70700}_{8200}$ & $0.517^{0.031}_{0.001}$ &$9.58^{92.88}_{7.29}$\\
TOI-789 & $23.2^{10.2}_{6.4}$ & $1.11^{0.37}_{0.29}$ & $1.29^{2.19}_{0.836}$ & $89.8^{42.6}_{42.4}$ & $0.045^{0.044}_{0.022}$ & $752^{503}_{395}$ & $0.505^{0.011}_{0.001}$ &$0.57^{0.74}_{0.45}$\\
TOI-451 & $4.07^{0.9}_{0.9}$ & $2.31^{0.8}_{0.59}$ & $6.53^{3.92}_{2.43}$ & $88.4^{2.8}_{2.2}$ & $0.045^{0.043}_{0.022}$ & $579^{403}_{298}$ & $0.907^{0.089}_{0.001}$ &$2.71^{2.06}_{1.13}$\\
TOI-487 & $50.7^{25.2}_{11.6}$ & $2.43^{0.84}_{0.63}$ & $7.08^{4.24}_{2.67}$ & $90.1^{43.1}_{43.0}$ & $0.07^{0.063}_{0.033}$ & $488^{1030}_{1010}$ & $0.51^{0.04}_{0.009}$ &$1.23^{0.46}_{0.77}$\\
HR 858 & $25.5^{9.2}_{5.5}$ & $2.05^{0.7}_{0.53}$ & $5.41^{3.19}_{2.04}$ & $86.5^{2.8}_{2.2}$ & $0.044^{0.043}_{0.021}$ & $206^{141}_{107}$ & $0.325^{0.517}_{0.001}$ &$1.07^{0.86}_{0.49}$\\
TOI-125 & $40.3^{14.8}_{7.8}$ & $2.79^{0.96}_{0.71}$ & $8.8^{5.24}_{3.27}$ & $89.9^{2.2}_{2.8}$ & $0.044^{0.043}_{0.021}$ & $906^{624}_{461}$ & $0.555^{0.318}_{0.001}$ &$1.82^{1.39}_{0.82}$\\
\enddata
\end{deluxetable*}
\end{longrotatetable}

%% file: packing_table_clust.tex
\begin{deluxetable}{C|RCCC|C}
\tablecolumns{6}
\tablecaption{Fraction of the 2024 sample of TESS multi-planet systems for which the corresponding additional predicted planet under the clustered period model (PCM) would be dynamically allowed for given critical separations ($\Delta_{crit})$, split by planet multiplicity ($N_{pl}$). The number of systems per planet multiplicity bin ($N_{sys}$) are also shown.}
\label{tab:packingclust}
\tablehead{
    \colhead{$N_{pl}$} &
    \multicolumn{4}{c}{Fraction of predicted additional planets allowed} &
    \colhead{$N_{sys}$}} 
\startdata
 & $\Delta_{crit} = 8$ & $\Delta_{crit} = 10$ & $\Delta_{crit} = 12.3$ & $\Delta_{crit} = 21.7$ \\
 \hline
2     &  0.96 &  0.95 &  0.86 &  0.03 &        128 \\
3     &  0.89 &  0.89 &  0.64 &  0.11 &         36 \\
4     &  1.00 &  0.91 &  0.82 &  0.00 &         11 \\
5     &  1.00 &  1.00 &  1.00 &  0.33 &          3 \\
6     &  1.00 &  0.75 &  0.75 &  0.00 &          4 \\
\hline
Total &  0.95 &  0.87 &  0.60 &  0.03 &  183 \\
\enddata
\end{deluxetable}

%% file: packing_table_ratio.tex
\begin{deluxetable}{L|CCCC|C}
\tablecolumns{6}
\tablecaption{Fraction of the 2024 sample of TESS multi-planet systems for which the corresponding additional predicted planet under the period ratio model (PRM) would be dynamically allowed for given critical separations ($\Delta_{crit})$, split by planet multiplicity ($N_{pl}$). The number of systems per planet multiplicity bin ($N_{sys}$) are also shown.}
\label{tab:packingratio}
\tablehead{
    \colhead{$N_{pl}$} &
    \multicolumn{4}{c}{Fraction of predicted additional planets allowed} &
    \colhead{$N_{sys}$}} 
\startdata
 & $\Delta_{crit} = 8$ & $\Delta_{crit} = 10$ & $\Delta_{crit} = 12.3$ & $\Delta_{crit} = 21.7$ \\
 \hline
2     &  0.96 &  0.95 &  0.89 &  0.07 &        128 \\
3     &  0.94 &  0.92 &  0.81 &  0.03 &         36 \\
4     &  1.00 &  1.00 &  0.91 &  0.00 &         11 \\
5     &  1.00 &  1.00 &  1.00 &  0.00 &          3 \\
6     &  1.00 &  1.00 &  0.75 &  0.00 &          4 \\
\hline
Total &  0.96 &  0.93 &  0.85 &  0.07 &  183 \\
\enddata
\end{deluxetable}

%% file: packing_table_nopred.tex
\begin{deluxetable}{L|CCCC|C}
\tablecolumns{6}
\tablecaption{Fraction of the 2024 sample of TESS multi-planet systems for which an additional $1 M_\oplus$ planet would be dynamically allowable (the ``unpacking fraction'') for given critical separations ($\Delta_{crit})$, split by planet multiplicity ($N_{pl}$). The number of systems per planet multiplicity bin ($N_{sys}$) are also shown.}
\label{tab:packing}
\tablehead{
    \colhead{$N_{pl}$} &
    \multicolumn{4}{c}{Fraction of predicted additional planets allowed} &
    \colhead{$N_{sys}$}} 
\startdata
 & $\Delta_{crit} = 8$ & $\Delta_{crit} = 10$ & $\Delta_{crit} = 12.3$ & $\Delta_{crit} = 21.7$ \\
 \hline
2     &  1.00 &  1.00 &  1.00 &  0.07 &        128 \\
3     &  0.50 &  0.50 &  0.00 &  0.00 &         36 \\
4     &  0.83 &  0.50 &  0.33 &  0.00 &         11 \\
5     &  0.75 &  0.25 &  0.25 &  0.00 &          3 \\
6     &  0.60 &  0.30 &  0.10 &  0.00 &          4 \\
\hline
Total &  0.73 &  0.61 &  0.47 &  0.21 &  183 \\
\enddata
\end{deluxetable}

%% file: main.bbl
\begin{thebibliography}{}
\expandafter\ifx\csname natexlab\endcsname\relax\def\natexlab#1{#1}\fi
\providecommand{\url}[1]{\href{#1}{#1}}
\providecommand{\dodoi}[1]{doi:~\href{http://doi.org/#1}{\nolinkurl{#1}}}
\providecommand{\doeprint}[1]{\href{http://ascl.net/#1}{\nolinkurl{http://ascl.net/#1}}}
\providecommand{\doarXiv}[1]{\href{https://arxiv.org/abs/#1}{\nolinkurl{https://arxiv.org/abs/#1}}}

\bibitem[{{Adams} {et~al.}(2020){Adams}, {Batygin}, {Bloch}, \&
  {Laughlin}}]{adams+20}
{Adams}, F.~C., {Batygin}, K., {Bloch}, A.~M., \& {Laughlin}, G. 2020, \mnras,
  493, 5520, \dodoi{10.1093/mnras/staa624}

\bibitem[{{Aigrain} {et~al.}(2015){Aigrain}, {Hodgkin}, {Irwin}, {Lewis}, \&
  {Roberts}}]{aigrain+15}
{Aigrain}, S., {Hodgkin}, S.~T., {Irwin}, M.~J., {Lewis}, J.~R., \& {Roberts},
  S.~J. 2015, \mnras, 447, 2880, \dodoi{10.1093/mnras/stu2638}

\bibitem[{{Astropy Collaboration} {et~al.}(2013){Astropy Collaboration},
  {Robitaille}, {Tollerud}, {Greenfield}, {Droettboom}, {Bray}, {Aldcroft},
  {Davis}, {Ginsburg}, {Price-Whelan}, {Kerzendorf}, {Conley}, {Crighton},
  {Barbary}, {Muna}, {Ferguson}, {Grollier}, {Parikh}, {Nair}, {Unther},
  {Deil}, {Woillez}, {Conseil}, {Kramer}, {Turner}, {Singer}, {Fox}, {Weaver},
  {Zabalza}, {Edwards}, {Azalee Bostroem}, {Burke}, {Casey}, {Crawford},
  {Dencheva}, {Ely}, {Jenness}, {Labrie}, {Lim}, {Pierfederici}, {Pontzen},
  {Ptak}, {Refsdal}, {Servillat}, \& {Streicher}}]{astropy:2013}
{Astropy Collaboration}, {Robitaille}, T.~P., {Tollerud}, E.~J., {et~al.} 2013,
  \aap, 558, A33, \dodoi{10.1051/0004-6361/201322068}

\bibitem[{{Astropy Collaboration} {et~al.}(2018){Astropy Collaboration},
  {Price-Whelan}, {Sip{\H{o}}cz}, {G{\"u}nther}, {Lim}, {Crawford}, {Conseil},
  {Shupe}, {Craig}, {Dencheva}, {Ginsburg}, {Vand erPlas}, {Bradley},
  {P{\'e}rez-Su{\'a}rez}, {de Val-Borro}, {Aldcroft}, {Cruz}, {Robitaille},
  {Tollerud}, {Ardelean}, {Babej}, {Bach}, {Bachetti}, {Bakanov}, {Bamford},
  {Barentsen}, {Barmby}, {Baumbach}, {Berry}, {Biscani}, {Boquien}, {Bostroem},
  {Bouma}, {Brammer}, {Bray}, {Breytenbach}, {Buddelmeijer}, {Burke},
  {Calderone}, {Cano Rodr{\'\i}guez}, {Cara}, {Cardoso}, {Cheedella}, {Copin},
  {Corrales}, {Crichton}, {D'Avella}, {Deil}, {Depagne}, {Dietrich}, {Donath},
  {Droettboom}, {Earl}, {Erben}, {Fabbro}, {Ferreira}, {Finethy}, {Fox},
  {Garrison}, {Gibbons}, {Goldstein}, {Gommers}, {Greco}, {Greenfield},
  {Groener}, {Grollier}, {Hagen}, {Hirst}, {Homeier}, {Horton}, {Hosseinzadeh},
  {Hu}, {Hunkeler}, {Ivezi{\'c}}, {Jain}, {Jenness}, {Kanarek}, {Kendrew},
  {Kern}, {Kerzendorf}, {Khvalko}, {King}, {Kirkby}, {Kulkarni}, {Kumar},
  {Lee}, {Lenz}, {Littlefair}, {Ma}, {Macleod}, {Mastropietro}, {McCully},
  {Montagnac}, {Morris}, {Mueller}, {Mumford}, {Muna}, {Murphy}, {Nelson},
  {Nguyen}, {Ninan}, {N{\"o}the}, {Ogaz}, {Oh}, {Parejko}, {Parley}, {Pascual},
  {Patil}, {Patil}, {Plunkett}, {Prochaska}, {Rastogi}, {Reddy Janga},
  {Sabater}, {Sakurikar}, {Seifert}, {Sherbert}, {Sherwood-Taylor}, {Shih},
  {Sick}, {Silbiger}, {Singanamalla}, {Singer}, {Sladen}, {Sooley},
  {Sornarajah}, {Streicher}, {Teuben}, {Thomas}, {Tremblay}, {Turner},
  {Terr{\'o}n}, {van Kerkwijk}, {de la Vega}, {Watkins}, {Weaver}, {Whitmore},
  {Woillez}, {Zabalza}, \& {Astropy Contributors}}]{astropy:2018}
{Astropy Collaboration}, {Price-Whelan}, A.~M., {Sip{\H{o}}cz}, B.~M., {et~al.}
  2018, \aj, 156, 123, \dodoi{10.3847/1538-3881/aabc4f}

\bibitem[{{Astropy Collaboration} {et~al.}(2022){Astropy Collaboration},
  {Price-Whelan}, {Lim}, {Earl}, {Starkman}, {Bradley}, {Shupe}, {Patil},
  {Corrales}, {Brasseur}, {N{"o}the}, {Donath}, {Tollerud}, {Morris},
  {Ginsburg}, {Vaher}, {Weaver}, {Tocknell}, {Jamieson}, {van Kerkwijk},
  {Robitaille}, {Merry}, {Bachetti}, {G{"u}nther}, {Aldcroft},
  {Alvarado-Montes}, {Archibald}, {B{'o}di}, {Bapat}, {Barentsen}, {Baz{'a}n},
  {Biswas}, {Boquien}, {Burke}, {Cara}, {Cara}, {Conroy}, {Conseil}, {Craig},
  {Cross}, {Cruz}, {D'Eugenio}, {Dencheva}, {Devillepoix}, {Dietrich},
  {Eigenbrot}, {Erben}, {Ferreira}, {Foreman-Mackey}, {Fox}, {Freij}, {Garg},
  {Geda}, {Glattly}, {Gondhalekar}, {Gordon}, {Grant}, {Greenfield}, {Groener},
  {Guest}, {Gurovich}, {Handberg}, {Hart}, {Hatfield-Dodds}, {Homeier},
  {Hosseinzadeh}, {Jenness}, {Jones}, {Joseph}, {Kalmbach}, {Karamehmetoglu},
  {Ka{l}uszy{'n}ski}, {Kelley}, {Kern}, {Kerzendorf}, {Koch}, {Kulumani},
  {Lee}, {Ly}, {Ma}, {MacBride}, {Maljaars}, {Muna}, {Murphy}, {Norman},
  {O'Steen}, {Oman}, {Pacifici}, {Pascual}, {Pascual-Granado}, {Patil},
  {Perren}, {Pickering}, {Rastogi}, {Roulston}, {Ryan}, {Rykoff}, {Sabater},
  {Sakurikar}, {Salgado}, {Sanghi}, {Saunders}, {Savchenko}, {Schwardt},
  {Seifert-Eckert}, {Shih}, {Jain}, {Shukla}, {Sick}, {Simpson},
  {Singanamalla}, {Singer}, {Singhal}, {Sinha}, {Sip{H{o}}cz}, {Spitler},
  {Stansby}, {Streicher}, {{{S}}umak}, {Swinbank}, {Taranu}, {Tewary},
  {Tremblay}, {Val-Borro}, {Van Kooten}, {Vasovi{'c}}, {Verma}, {de Miranda
  Cardoso}, {Williams}, {Wilson}, {Winkel}, {Wood-Vasey}, {Xue}, {Yoachim},
  {Zhang}, {Zonca}, \& {Astropy Project Contributors}}]{astropy:2022}
{Astropy Collaboration}, {Price-Whelan}, A.~M., {Lim}, P.~L., {et~al.} 2022,
  \apj, 935, 167, \dodoi{10.3847/1538-4357/ac7c74}

\bibitem[{{Badenas-Agusti} {et~al.}(2020){Badenas-Agusti}, {G{\"u}nther},
  {Daylan}, {Mikal-Evans}, {Vanderburg}, {Huang}, {Matthews}, {Rackham},
  {Bieryla}, {Stassun}, {Kane}, {Shporer}, {Fulton}, {Hill}, {Nowak}, {Ribas},
  {Pall{\'e}}, {Jenkins}, {Latham}, {Seager}, {Ricker}, {Vanderspek}, {Winn},
  {Abril-Pla}, {Collins}, {Serra}, {Niraula}, {Rustamkulov}, {Barclay},
  {Crossfield}, {Howell}, {Ciardi}, {Gonzales}, {Schlieder}, {Caldwell},
  {Fausnaugh}, {McDermott}, {Paegert}, {Pepper}, {Rose}, \&
  {Twicken}}]{badenas-agusti+2020}
{Badenas-Agusti}, M., {G{\"u}nther}, M.~N., {Daylan}, T., {et~al.} 2020, \aj,
  160, 113, \dodoi{10.3847/1538-3881/aba0b5}

\bibitem[{{Barclay} {et~al.}(2017){Barclay}, {Quintana}, {Raymond}, \&
  {Penny}}]{barclay+17}
{Barclay}, T., {Quintana}, E.~V., {Raymond}, S.~N., \& {Penny}, M.~T. 2017,
  \apj, 841, 86, \dodoi{10.3847/1538-4357/aa705b}

\bibitem[{{Barnes} \& {Raymond}(2004)}]{barnes+raymond04}
{Barnes}, R., \& {Raymond}, S.~N. 2004, \apj, 617, 569, \dodoi{10.1086/423419}

\bibitem[{{Batalha}(2014)}]{batalha14}
{Batalha}, N.~M. 2014, Proceedings of the National Academy of Science, 111,
  12647, \dodoi{10.1073/pnas.1304196111}

\bibitem[{{Bennett} \& {Rhie}(2002)}]{bennett+rhie02}
{Bennett}, D.~P., \& {Rhie}, S.~H. 2002, \apj, 574, 985, \dodoi{10.1086/340977}

\bibitem[{{Berger} {et~al.}(2020){Berger}, {Huber}, {van Saders}, {Gaidos},
  {Tayar}, \& {Kraus}}]{berger+20}
{Berger}, T.~A., {Huber}, D., {van Saders}, J.~L., {et~al.} 2020, \aj, 159,
  280, \dodoi{10.3847/1538-3881/159/6/280}

\bibitem[{{Brewer} {et~al.}(2023){Brewer}, {Zhao}, {Fischer}, {Roettenbacher},
  {Henry}, {Llama}, {Szymkowiak}, {Cabot}, {Weiss}, \& {McCarthy}}]{brewer+23}
{Brewer}, J.~M., {Zhao}, L.~L., {Fischer}, D.~A., {et~al.} 2023, \aj, 166, 46,
  \dodoi{10.3847/1538-3881/acdd6f}

\bibitem[{{Carrera} {et~al.}(2018){Carrera}, {Ford}, {Izidoro},
  {Jontof-Hutter}, {Raymond}, \& {Wolfgang}}]{Carrera2018}
{Carrera}, D., {Ford}, E.~B., {Izidoro}, A., {et~al.} 2018, ApJ, 866, 104,
  \dodoi{10.3847/1538-4357/aadf8a}

\bibitem[{{Chatterjee} {et~al.}(2008){Chatterjee}, {Ford}, {Matsumura}, \&
  {Rasio}}]{Chatterjee2008}
{Chatterjee}, S., {Ford}, E.~B., {Matsumura}, S., \& {Rasio}, F.~A. 2008, ApJ,
  686, 580, \dodoi{10.1086/590227}

\bibitem[{{Chauvin}(2024)}]{chauvin23}
{Chauvin}, G. 2024, Comptes Rendus Physique, 24, 139,
  \dodoi{10.5802/crphys.139}

\bibitem[{{Chen} \& {Kipping}(2017)}]{chen&kipping17}
{Chen}, J., \& {Kipping}, D. 2017, \apj, 834, 17,
  \dodoi{10.3847/1538-4357/834/1/17}

\bibitem[{{Choksi} \& {Chiang}(2020)}]{choksi+chiang20}
{Choksi}, N., \& {Chiang}, E. 2020, \mnras, 495, 4192,
  \dodoi{10.1093/mnras/staa1421}

\bibitem[{{Dai} {et~al.}(2018){Dai}, {Masuda}, \& {Winn}}]{Dai2018}
{Dai}, F., {Masuda}, K., \& {Winn}, J.~N. 2018, ApJL, 864, L38,
  \dodoi{10.3847/2041-8213/aadd4f}

\bibitem[{{Dawson} {et~al.}(2016){Dawson}, {Lee}, \& {Chiang}}]{Dawson2016}
{Dawson}, R.~I., {Lee}, E.~J., \& {Chiang}, E. 2016, ApJ, 822, 54,
  \dodoi{10.3847/0004-637X/822/1/54}

\bibitem[{{Denham} {et~al.}(2019){Denham}, {Naoz}, {Hoang}, {Stephan}, \&
  {Farr}}]{denham+19}
{Denham}, P., {Naoz}, S., {Hoang}, B.-M., {Stephan}, A.~P., \& {Farr}, W.~M.
  2019, \mnras, 482, 4146, \dodoi{10.1093/mnras/sty2830}

\bibitem[{{Dietrich} \& {Apai}(2020)}]{dyn2020}
{Dietrich}, J., \& {Apai}, D. 2020, \aj, 160, 107,
  \dodoi{10.3847/1538-3881/aba61d}

\bibitem[{{Dietrich} {et~al.}(2022){Dietrich}, {Apai}, \&
  {Malhotra}}]{dietrich+22}
{Dietrich}, J., {Apai}, D., \& {Malhotra}, R. 2022, \aj, 163, 88,
  \dodoi{10.3847/1538-3881/ac4166}

\bibitem[{{Dietrich} {et~al.}(2024){Dietrich}, {Malhotra}, \&
  {Apai}}]{Dietrich2024}
{Dietrich}, J., {Malhotra}, R., \& {Apai}, D. 2024, AJ, 167, 46,
  \dodoi{10.3847/1538-3881/ad1244}

\bibitem[{{Dong} \& {Zhu}(2013)}]{dong&zhu13}
{Dong}, S., \& {Zhu}, Z. 2013, \apj, 778, 53,
  \dodoi{10.1088/0004-637X/778/1/53}

\bibitem[{{Dransfield} {et~al.}(2022){Dransfield}, {Triaud}, {Guillot},
  {Mekarnia}, {Nesvorn{\'y}}, {Crouzet}, {Abe}, {Agabi}, {Buttu}, {Cabrera},
  {Gandolfi}, {G{\"u}nther}, {Rodler}, {Schmider}, {Stee}, {Suarez}, {Collins},
  {D{\'e}vora-Pajares}, {Howell}, {Matthews}, {Standing}, {Stassun},
  {Stockdale}, {Quinn}, {Ziegler}, {Crossfield}, {Lissauer}, {Mann}, {Matson},
  {Schlieder}, \& {Zhou}}]{dransfield+2022}
{Dransfield}, G., {Triaud}, A. H.~M.~J., {Guillot}, T., {et~al.} 2022, \mnras,
  515, 1328, \dodoi{10.1093/mnras/stac1383}

\bibitem[{{Dr{\k{a}}{\.z}kowska} {et~al.}(2023){Dr{\k{a}}{\.z}kowska},
  {Bitsch}, {Lambrechts}, {Mulders}, {Harsono}, {Vazan}, {Liu}, {Ormel},
  {Kretke}, \& {Morbidelli}}]{drazkowska+23}
{Dr{\k{a}}{\.z}kowska}, J., {Bitsch}, B., {Lambrechts}, M., {et~al.} 2023, in
  Astronomical Society of the Pacific Conference Series, Vol. 534, Protostars
  and Planets VII, ed. S.~{Inutsuka}, Y.~{Aikawa}, T.~{Muto}, K.~{Tomida}, \&
  M.~{Tamura}, 717, \dodoi{10.48550/arXiv.2203.09759}

\bibitem[{{Fabrycky} \& {Tremaine}(2007)}]{fabrycky+tremaine07}
{Fabrycky}, D., \& {Tremaine}, S. 2007, \apj, 669, 1298, \dodoi{10.1086/521702}

\bibitem[{{Fabrycky} {et~al.}(2014){Fabrycky}, {Lissauer}, {Ragozzine}, {Rowe},
  {Steffen}, {Agol}, {Barclay}, {Batalha}, {Borucki}, {Ciardi}, {Ford},
  {Gautier}, {Geary}, {Holman}, {Jenkins}, {Li}, {Morehead}, {Morris},
  {Shporer}, {Smith}, {Still}, \& {Van Cleve}}]{fabrycky+2014}
{Fabrycky}, D.~C., {Lissauer}, J.~J., {Ragozzine}, D., {et~al.} 2014, \apj,
  790, 146, \dodoi{10.1088/0004-637X/790/2/146}

\bibitem[{{Fang} \& {Margot}(2013)}]{fang+13}
{Fang}, J., \& {Margot}, J.-L. 2013, \apj, 767, 115,
  \dodoi{10.1088/0004-637X/767/2/115}

\bibitem[{{Fischer} \& {Valenti}(2005)}]{fischer+05}
{Fischer}, D.~A., \& {Valenti}, J. 2005, \apj, 622, 1102,
  \dodoi{10.1086/428383}

\bibitem[{{Fulton} \& {Petigura}(2018)}]{fulton+18}
{Fulton}, B.~J., \& {Petigura}, E.~A. 2018, \aj, 156, 264,
  \dodoi{10.3847/1538-3881/aae828}

\bibitem[{{Ghosh} \& {Chatterjee}(2024)}]{ghosh+chatterjee24}
{Ghosh}, T., \& {Chatterjee}, S. 2024, \mnras, 527, 79,
  \dodoi{10.1093/mnras/stad2962}

\bibitem[{{Goldberg} \& {Batygin}(2022)}]{Goldberg2022}
{Goldberg}, M., \& {Batygin}, K. 2022, AJ, 163, 201,
  \dodoi{10.3847/1538-3881/ac5961}

\bibitem[{{Gondhalekar} {et~al.}(2023){Gondhalekar}, {Feigelson}, {Caceres},
  {Montalto}, \& {Saha}}]{gondhalekar+23}
{Gondhalekar}, Y., {Feigelson}, E.~D., {Caceres}, G.~A., {Montalto}, M., \&
  {Saha}, S. 2023, \apjl, 959, L16, \dodoi{10.3847/2041-8213/ad0844}

\bibitem[{{Han} \& {Brandt}(2023)}]{han+23}
{Han}, T., \& {Brandt}, T.~D. 2023, \aj, 165, 71,
  \dodoi{10.3847/1538-3881/acaaa7}

\bibitem[{{He} {et~al.}(2019){He}, {Ford}, \& {Ragozzine}}]{he+2019}
{He}, M.~Y., {Ford}, E.~B., \& {Ragozzine}, D. 2019, \mnras, 490, 4575,
  \dodoi{10.1093/mnras/stz2869}

\bibitem[{{He} {et~al.}(2020){He}, {Ford}, {Ragozzine}, \& {Carrera}}]{he+2020}
{He}, M.~Y., {Ford}, E.~B., {Ragozzine}, D., \& {Carrera}, D. 2020, \aj, 160,
  276, \dodoi{10.3847/1538-3881/abba18}

\bibitem[{{Hippke} {et~al.}(2019){Hippke}, {David}, {Mulders}, \&
  {Heller}}]{wotan}
{Hippke}, M., {David}, T.~J., {Mulders}, G.~D., \& {Heller}, R. 2019, \aj, 158,
  143, \dodoi{10.3847/1538-3881/ab3984}

\bibitem[{{Huang} {et~al.}(2020){Huang}, {Vanderburg}, {P{\'a}l}, {Sha}, {Yu},
  {Fong}, {Fausnaugh}, {Shporer}, {Guerrero}, {Vanderspek}, \& {Ricker}}]{qlp}
{Huang}, C.~X., {Vanderburg}, A., {P{\'a}l}, A., {et~al.} 2020, Research Notes
  of the American Astronomical Society, 4, 204,
  \dodoi{10.3847/2515-5172/abca2e}

\bibitem[{{Izidoro} {et~al.}(2021){Izidoro}, {Bitsch}, {Raymond}, {Johansen},
  {Morbidelli}, {Lambrechts}, \& {Jacobson}}]{izidoro+21}
{Izidoro}, A., {Bitsch}, B., {Raymond}, S.~N., {et~al.} 2021, \aap, 650, A152,
  \dodoi{10.1051/0004-6361/201935336}

\bibitem[{{Izidoro} {et~al.}(2017){Izidoro}, {Ogihara}, {Raymond},
  {Morbidelli}, {Pierens}, {Bitsch}, {Cossou}, \& {Hersant}}]{Izidoro2017}
{Izidoro}, A., {Ogihara}, M., {Raymond}, S.~N., {et~al.} 2017, MNRAS, 470,
  1750, \dodoi{10.1093/mnras/stx1232}

\bibitem[{{Jackson} {et~al.}(2008){Jackson}, {Greenberg}, \&
  {Barnes}}]{jackson+08}
{Jackson}, B., {Greenberg}, R., \& {Barnes}, R. 2008, \apj, 678, 1396,
  \dodoi{10.1086/529187}

\bibitem[{{Jenkins} {et~al.}(2016){Jenkins}, {Twicken}, {McCauliff},
  {Campbell}, {Sanderfer}, {Lung}, {Mansouri-Samani}, {Girouard}, {Tenenbaum},
  {Klaus}, {Smith}, {Caldwell}, {Chacon}, {Henze}, {Heiges}, {Latham},
  {Morgan}, {Swade}, {Rinehart}, \& {Vanderspek}}]{spoc}
{Jenkins}, J.~M., {Twicken}, J.~D., {McCauliff}, S., {et~al.} 2016, in
  \procspie, Vol. 9913, Software and Cyberinfrastructure for Astronomy IV,
  99133E, \dodoi{10.1117/12.2233418}

\bibitem[{{Juri{\'c}} \& {Tremaine}(2008)}]{Juric2008}
{Juri{\'c}}, M., \& {Tremaine}, S. 2008, ApJ, 686, 603, \dodoi{10.1086/590047}

\bibitem[{{Kov{\'a}cs} {et~al.}(2002){Kov{\'a}cs}, {Zucker}, \& {Mazeh}}]{bls}
{Kov{\'a}cs}, G., {Zucker}, S., \& {Mazeh}, T. 2002, \aap, 391, 369,
  \dodoi{10.1051/0004-6361:20020802}

\bibitem[{{Kozai}(1962)}]{kozai62}
{Kozai}, Y. 1962, \aj, 67, 591, \dodoi{10.1086/108790}

\bibitem[{{Kreidberg}(2015)}]{batman}
{Kreidberg}, L. 2015, \pasp, 127, 1161, \dodoi{10.1086/683602}

\bibitem[{{Lidov}(1962)}]{lidov62}
{Lidov}, M.~L. 1962, \planss, 9, 719, \dodoi{10.1016/0032-0633(62)90129-0}

\bibitem[{{Lightkurve Collaboration} {et~al.}(2018){Lightkurve Collaboration},
  {Cardoso}, {Hedges}, {Gully-Santiago}, {Saunders}, {Cody}, {Barclay}, {Hall},
  {Sagear}, {Turtelboom}, {Zhang}, {Tzanidakis}, {Mighell}, {Coughlin}, {Bell},
  {Berta-Thompson}, {Williams}, {Dotson}, \& {Barentsen}}]{lk}
{Lightkurve Collaboration}, {Cardoso}, J.~V.~d.~M., {Hedges}, C., {et~al.}
  2018, {Lightkurve: Kepler and TESS time series analysis in Python},
  Astrophysics Source Code Library.
\newblock \doeprint{1812.013}

\bibitem[{{Lissauer} {et~al.}(2011){Lissauer}, {Ragozzine}, {Fabrycky},
  {Steffen}, {Ford}, {Jenkins}, {Shporer}, {Holman}, {Rowe}, {Quintana},
  {Batalha}, {Borucki}, {Bryson}, {Caldwell}, {Carter}, {Ciardi}, {Dunham},
  {Fortney}, {Gautier}, {Howell}, {Koch}, {Latham}, {Marcy}, {Morehead}, \&
  {Sasselov}}]{lissauer+2011}
{Lissauer}, J.~J., {Ragozzine}, D., {Fabrycky}, D.~C., {et~al.} 2011, \apjs,
  197, 8, \dodoi{10.1088/0067-0049/197/1/8}

\bibitem[{{Lithwick} \& {Wu}(2012)}]{lithwick+12a}
{Lithwick}, Y., \& {Wu}, Y. 2012, \apjl, 756, L11,
  \dodoi{10.1088/2041-8205/756/1/L11}

\bibitem[{{Lithwick} {et~al.}(2012){Lithwick}, {Xie}, \& {Wu}}]{lithwick+12b}
{Lithwick}, Y., {Xie}, J., \& {Wu}, Y. 2012, \apj, 761, 122,
  \dodoi{10.1088/0004-637X/761/2/122}

\bibitem[{{Lubin} {et~al.}(2022){Lubin}, {Van Zandt}, {Holcomb}, {Weiss},
  {Petigura}, {Robertson}, {Akana Murphy}, {Scarsdale}, {Batygin}, {Polanski},
  {Batalha}, {Crossfield}, {Dressing}, {Fulton}, {Howard}, {Huber}, {Isaacson},
  {Kane}, {Roy}, {Beard}, {Blunt}, {Chontos}, {Dai}, {Dalba}, {Gary},
  {Giacalone}, {Hill}, {Mayo}, {Mo{\v{c}}nik}, {Kosiarek}, {Rice}, {Rubenzahl},
  {Latham}, {Seager}, {Winn}, \& {Gary}}]{lubin+22}
{Lubin}, J., {Van Zandt}, J., {Holcomb}, R., {et~al.} 2022, \aj, 163, 101,
  \dodoi{10.3847/1538-3881/ac3d38}

\bibitem[{{Lubin} {et~al.}(2024){Lubin}, {Petigura}, {Van Zandt}, {Beard},
  {Dai}, {Halverson}, {Holcomb}, {Howard}, {Isaacson}, {Luhn}, {Robertson},
  {Rubenzahl}, {Stef{\'a}nsson}, {Winn}, {Brodheim}, {Deich}, {Hill}, {Gibson},
  {Holden}, {Householder}, {Laher}, {Lanclos}, {Payne}, {Roy}, {Smith},
  {Shaum}, {Schwab}, \& {Walawender}}]{lubin+24}
{Lubin}, J., {Petigura}, E.~A., {Van Zandt}, J., {et~al.} 2024, \aj, 168, 196,
  \dodoi{10.3847/1538-3881/ad79ed}

\bibitem[{{Luque} \& {Pall{\'e}}(2022)}]{luque+palle22}
{Luque}, R., \& {Pall{\'e}}, E. 2022, Science, 377, 1211,
  \dodoi{10.1126/science.abl7164}

\bibitem[{{MacDonald} {et~al.}(2020){MacDonald}, {Dawson}, {Morrison}, {Lee},
  \& {Khandelwal}}]{macdonald+20}
{MacDonald}, M.~G., {Dawson}, R.~I., {Morrison}, S.~J., {Lee}, E.~J., \&
  {Khandelwal}, A. 2020, \apj, 891, 20, \dodoi{10.3847/1538-4357/ab6f04}

\bibitem[{{Millholland} {et~al.}(2017){Millholland}, {Wang}, \&
  {Laughlin}}]{millholland+2017}
{Millholland}, S., {Wang}, S., \& {Laughlin}, G. 2017, \apjl, 849, L33,
  \dodoi{10.3847/2041-8213/aa9714}

\bibitem[{{Millholland} {et~al.}(2022){Millholland}, {He}, \&
  {Zink}}]{millholland2022}
{Millholland}, S.~C., {He}, M.~Y., \& {Zink}, J.~K. 2022, \aj, 164, 72,
  \dodoi{10.3847/1538-3881/ac7c67}

\bibitem[{{Mills} {et~al.}(2019){Mills}, {Howard}, {Petigura}, {Fulton},
  {Isaacson}, \& {Weiss}}]{Mills+2019}
{Mills}, S.~M., {Howard}, A.~W., {Petigura}, E.~A., {et~al.} 2019, \aj, 157,
  198, \dodoi{10.3847/1538-3881/ab1009}

\bibitem[{{Montet} {et~al.}(2017){Montet}, {Yee}, \& {Penny}}]{montet+17}
{Montet}, B.~T., {Yee}, J.~C., \& {Penny}, M.~T. 2017, \pasp, 129, 044401,
  \dodoi{10.1088/1538-3873/aa57fb}

\bibitem[{{Mordasini}(2018)}]{mordasini18}
{Mordasini}, C. 2018, in Handbook of Exoplanets, ed. H.~J. {Deeg} \& J.~A.
  {Belmonte} (Springer), 143, \dodoi{10.1007/978-3-319-55333-7_143}

\bibitem[{{Mulders} {et~al.}(2015){Mulders}, {Pascucci}, \&
  {Apai}}]{Mulders2015a}
{Mulders}, G.~D., {Pascucci}, I., \& {Apai}, D. 2015, ApJ, 798, 112,
  \dodoi{10.1088/0004-637X/798/2/112}

\bibitem[{{Mulders} {et~al.}(2018){Mulders}, {Pascucci}, {Apai}, \&
  {Ciesla}}]{mulders+2018}
{Mulders}, G.~D., {Pascucci}, I., {Apai}, D., \& {Ciesla}, F.~J. 2018, \aj,
  156, 24, \dodoi{10.3847/1538-3881/aac5ea}

\bibitem[{{NASA Exoplanet Archive}(2019)}]{nea1}
{NASA Exoplanet Archive}. 2019, Confirmed Planets Table,  IPAC,
  \dodoi{10.26133/NEA1}

\bibitem[{{NExScI}(2022)}]{exofop5}
{NExScI}. 2022, Exoplanet Follow-up Observing Program Web Service,  IPAC,
  \dodoi{10.26134/EXOFOP5}

\bibitem[{{Ning} {et~al.}(2018){Ning}, {Wolfgang}, \& {Ghosh}}]{ning+18}
{Ning}, B., {Wolfgang}, A., \& {Ghosh}, S. 2018, \apj, 869, 5,
  \dodoi{10.3847/1538-4357/aaeb31}

\bibitem[{{Obertas} {et~al.}(2023){Obertas}, {Tamayo}, \&
  {Murray}}]{obertas+23}
{Obertas}, A., {Tamayo}, D., \& {Murray}, N. 2023, \mnras, 526, 2118,
  \dodoi{10.1093/mnras/stad1921}

\bibitem[{{Ofir}(2014)}]{ofir14}
{Ofir}, A. 2014, \aap, 561, A138, \dodoi{10.1051/0004-6361/201220860}

\bibitem[{{Orell-Miquel} {et~al.}(2023){Orell-Miquel}, {Nowak}, {Murgas},
  {Palle}, {Morello}, {Luque}, {Badenas-Agusti}, {Ribas}, {Lafarga},
  {Espinoza}, {Morales}, {Zechmeister}, {Alqasim}, {Cochran}, {Gandolfi},
  {Goffo}, {Kab{\'a}th}, {Korth}, {Lam}, {Livingston}, {Muresan}, {Persson}, \&
  {Van Eylen}}]{orell-miquel+2023}
{Orell-Miquel}, J., {Nowak}, G., {Murgas}, F., {et~al.} 2023, \aap, 669, A40,
  \dodoi{10.1051/0004-6361/202244120}

\bibitem[{{Parviainen} {et~al.}(2024){Parviainen}, {Luque}, \&
  {Palle}}]{spright}
{Parviainen}, H., {Luque}, R., \& {Palle}, E. 2024, \mnras, 527, 5693,
  \dodoi{10.1093/mnras/stad3504}

\bibitem[{{Penny} {et~al.}(2019){Penny}, {Gaudi}, {Kerins}, {Rattenbury},
  {Mao}, {Robin}, \& {Calchi Novati}}]{penny+19}
{Penny}, M.~T., {Gaudi}, B.~S., {Kerins}, E., {et~al.} 2019, \apjs, 241, 3,
  \dodoi{10.3847/1538-4365/aafb69}

\bibitem[{{Pu} \& {Wu}(2015)}]{pu&wu15}
{Pu}, B., \& {Wu}, Y. 2015, \apj, 807, 44, \dodoi{10.1088/0004-637X/807/1/44}

\bibitem[{{Sandford} {et~al.}(2019){Sandford}, {Kipping}, \&
  {Collins}}]{sandford+2019}
{Sandford}, E., {Kipping}, D., \& {Collins}, M. 2019, \mnras, 489, 3162,
  \dodoi{10.1093/mnras/stz2350}

\bibitem[{Smirnov(1939)}]{smirnov1939estimation}
Smirnov, N.~V. 1939, Bull. Math. Univ. Moscou, 2, 3

\bibitem[{{Sobski} \& {Millholland}(2023)}]{sobski23}
{Sobski}, N., \& {Millholland}, S.~C. 2023, \apj, 954, 137,
  \dodoi{10.3847/1538-4357/ace966}

\bibitem[{{Southworth}(2011)}]{tepcat}
{Southworth}, J. 2011, \mnras, 417, 2166,
  \dodoi{10.1111/j.1365-2966.2011.19399.x}

\bibitem[{{Spergel} {et~al.}(2015){Spergel}, {Gehrels}, {Baltay}, {Bennett},
  {Breckinridge}, {Donahue}, {Dressler}, {Gaudi}, {Greene}, {Guyon}, {Hirata},
  {Kalirai}, {Kasdin}, {Macintosh}, {Moos}, {Perlmutter}, {Postman},
  {Rauscher}, {Rhodes}, {Wang}, {Weinberg}, {Benford}, {Hudson}, {Jeong},
  {Mellier}, {Traub}, {Yamada}, {Capak}, {Colbert}, {Masters}, {Penny},
  {Savransky}, {Stern}, {Zimmerman}, {Barry}, {Bartusek}, {Carpenter}, {Cheng},
  {Content}, {Dekens}, {Demers}, {Grady}, {Jackson}, {Kuan}, {Kruk}, {Melton},
  {Nemati}, {Parvin}, {Poberezhskiy}, {Peddie}, {Ruffa}, {Wallace}, {Whipple},
  {Wollack}, \& {Zhao}}]{spergel+15}
{Spergel}, D., {Gehrels}, N., {Baltay}, C., {et~al.} 2015, arXiv e-prints,
  arXiv:1503.03757, \dodoi{10.48550/arXiv.1503.03757}

\bibitem[{{Steffen} \& {Hwang}(2015)}]{steffen+2015}
{Steffen}, J.~H., \& {Hwang}, J.~A. 2015, \mnras, 448, 1956,
  \dodoi{10.1093/mnras/stv104}

\bibitem[{{STScI}(2018)}]{tic}
{STScI}. 2018, TESS Input Catalog and Candidate Target List,  STScI/MAST,
  \dodoi{10.17909/FWDT-2X66}

\bibitem[{{Sumi} {et~al.}(2011){Sumi}, {Kamiya}, {Bennett}, {Bond}, {Abe},
  {Botzler}, {Fukui}, {Furusawa}, {Hearnshaw}, {Itow}, {Kilmartin}, {Korpela},
  {Lin}, {Ling}, {Masuda}, {Matsubara}, {Miyake}, {Motomura}, {Muraki},
  {Nagaya}, {Nakamura}, {Ohnishi}, {Okumura}, {Perrott}, {Rattenbury}, {Saito},
  {Sako}, {Sullivan}, {Sweatman}, {Tristram}, {Udalski}, {Szyma{\'n}ski},
  {Kubiak}, {Pietrzy{\'n}ski}, {Poleski}, {Soszy{\'n}ski}, {Wyrzykowski},
  {Ulaczyk}, \& {Microlensing Observations in Astrophysics (MOA)
  Collaboration}}]{sumi+11}
{Sumi}, T., {Kamiya}, K., {Bennett}, D.~P., {et~al.} 2011, \nat, 473, 349,
  \dodoi{10.1038/nature10092}

\bibitem[{{Tamburo} {et~al.}(2023){Tamburo}, {Muirhead}, \&
  {Dressing}}]{tamburo+23}
{Tamburo}, P., {Muirhead}, P.~S., \& {Dressing}, C.~D. 2023, \aj, 165, 251,
  \dodoi{10.3847/1538-3881/acd1de}

\bibitem[{{Ti{\'o} Humphrey} \& {Quintana}(2020)}]{humphrey+2022}
{Ti{\'o} Humphrey}, A.~L., \& {Quintana}, E.~V. 2020, arXiv e-prints,
  arXiv:2011.03053, \dodoi{10.48550/arXiv.2011.03053}

\bibitem[{Turtelboom(2024)}]{zenodo}
Turtelboom, E.~V. 2024, Signal-to-Noise Ratios, DYNAMITE Posteriors, and
  Combined Detectability Plots for 173 TESS Multi-Planet Systems,  Zenodo,
  \dodoi{10.5281/ZENODO.14183270}

\bibitem[{{Tuson} {et~al.}(2023){Tuson}, {Queloz}, {Osborn}, {Wilson},
  {Hooton}, {Beck}, {Lendl}, {Olofsson}, {Fortier}, {Bonfanti}, {Brandeker},
  {Buchhave}, {Collier Cameron}, {Ciardi}, {Collins}, {Gandolfi}, {Garai},
  {Giacalone}, {Gomes da Silva}, {Howell}, {Patel}, {Persson}, {Serrano},
  {Sousa}, {Ulmer-Moll}, {Vanderburg}, {Ziegler}, {Alibert}, {Alonso},
  {Anglada}, {B{\'a}rczy}, {Barrado Navascues}, {Barros}, {Baumjohann}, {Beck},
  {Benz}, {Billot}, {Bonfils}, {Borsato}, {Broeg}, {Cabrera}, {Charnoz},
  {Conti}, {Csizmadia}, {Cubillos}, {Davies}, {Deleuil}, {Delrez}, {Demangeon},
  {Demory}, {Dragomir}, {Dressing}, {Ehrenreich}, {Erikson}, {Essack},
  {Farinato}, {Fossati}, {Fridlund}, {Furlan}, {Gill}, {Gillon}, {Gnilka},
  {Gonzales}, {G{\"u}del}, {G{\"u}nther}, {Hoyer}, {Isaak}, {Jenkins}, {Kiss},
  {Laskar}, {Latham}, {Law}, {Lecavelier des Etangs}, {Curto}, {Lovis},
  {Luque}, {Magrin}, {Mann}, {Maxted}, {Mayor}, {McDermott}, {Mecina},
  {Mordasini}, {Mortier}, {Nascimbeni}, {Ottensamer}, {Pagano}, {Pall{\'e}},
  {Peter}, {Piotto}, {Pollacco}, {Pritchard}, {Ragazzoni}, {Rando}, {Ratti},
  {Rauer}, {Ribas}, {Ricker}, {Rieder}, {Santos}, {Savel}, {Scandariato},
  {Schwarz}, {Seager}, {S{\'e}gransan}, {Shporer}, {Simon}, {Smith}, {Steller},
  {Stockdale}, {Szab{\'o}}, {Thomas}, {Torres}, {Tronsgaard}, {Udry}, {Ulmer},
  {Van Grootel}, {Vanderspek}, {Venturini}, {Walton}, {Winn}, \&
  {Wohler}}]{tuson+23}
{Tuson}, A., {Queloz}, D., {Osborn}, H.~P., {et~al.} 2023, \mnras, 523, 3090,
  \dodoi{10.1093/mnras/stad1369}

\bibitem[{{Ulmer-Moll} {et~al.}(2022){Ulmer-Moll}, {Lendl}, {Gill},
  {Villanueva}, {Hobson}, {Bouchy}, {Brahm}, {Dragomir}, {Grieves},
  {Mordasini}, {Anderson}, {Acton}, {Bayliss}, {Bieryla}, {Burleigh},
  {Casewell}, {Chaverot}, {Eigm{\"u}ller}, {Feliz}, {Gaudi}, {Gillen}, {Goad},
  {Gupta}, {G{\"u}nther}, {Henderson}, {Henning}, {Jenkins}, {Jones},
  {Jord{\'a}n}, {Kendall}, {Latham}, {Mireles}, {Moyano}, {Nadol}, {Osborn},
  {Pepper}, {Tala Pinto}, {Psaridi}, {Queloz}, {Quinn}, {Rojas}, {Sarkis},
  {Schlecker}, {Tilbrook}, {Torres}, {Trifonov}, {Udry}, {Vines}, {West},
  {Wheatley}, {Yao}, {Zhao}, \& {Zhou}}]{ulmer-moll+22}
{Ulmer-Moll}, S., {Lendl}, M., {Gill}, S., {et~al.} 2022, \aap, 666, A46,
  \dodoi{10.1051/0004-6361/202243583}

\bibitem[{{Weiss} {et~al.}(2018){Weiss}, {Marcy}, {Petigura}, {Fulton},
  {Howard}, {Winn}, {Isaacson}, {Morton}, {Hirsch}, {Sinukoff}, {Cumming},
  {Hebb}, \& {Cargile}}]{weiss+2018}
{Weiss}, L.~M., {Marcy}, G.~W., {Petigura}, E.~A., {et~al.} 2018, \aj, 155, 48,
  \dodoi{10.3847/1538-3881/aa9ff6}

\bibitem[{{Wilson} {et~al.}(2023){Wilson}, {Barclay}, {Powell}, {Schlieder},
  {Hedges}, {Montet}, {Quintana}, {Mcdonald}, {Penny}, {Espinoza}, \&
  {Kerins}}]{wilson+23}
{Wilson}, R.~F., {Barclay}, T., {Powell}, B.~P., {et~al.} 2023, \apjs, 269, 5,
  \dodoi{10.3847/1538-4365/acf3df}

\bibitem[{{Wu} {et~al.}(2024){Wu}, {Dewberry}, \& {Fuller}}]{wu+24}
{Wu}, S.~C., {Dewberry}, J.~W., \& {Fuller}, J. 2024, \apj, 963, 34,
  \dodoi{10.3847/1538-4357/ad1e54}

\bibitem[{{Xie} {et~al.}(2016){Xie}, {Dong}, {Zhu}, {Huber}, {Zheng}, {De Cat},
  {Fu}, {Liu}, {Luo}, {Wu}, {Zhang}, {Zhang}, {Zhou}, {Cao}, {Hou}, {Wang}, \&
  {Zhang}}]{xie+2016}
{Xie}, J.-W., {Dong}, S., {Zhu}, Z., {et~al.} 2016, Proceedings of the National
  Academy of Science, 113, 11431, \dodoi{10.1073/pnas.1604692113}

\bibitem[{{Yang} {et~al.}(2020){Yang}, {Xie}, \& {Zhou}}]{yang+20}
{Yang}, J.-Y., {Xie}, J.-W., \& {Zhou}, J.-L. 2020, \aj, 159, 164,
  \dodoi{10.3847/1538-3881/ab7373}

\bibitem[{{Zawadzki} {et~al.}(2022){Zawadzki}, {Carrera}, \&
  {Ford}}]{Zawadzki2022}
{Zawadzki}, B., {Carrera}, D., \& {Ford}, E.~B. 2022, ApJ, 937, 53,
  \dodoi{10.3847/1538-4357/ac8b04}

\bibitem[{{Zhu} {et~al.}(2018){Zhu}, {Petrovich}, {Wu}, {Dong}, \&
  {Xie}}]{zhu+18}
{Zhu}, W., {Petrovich}, C., {Wu}, Y., {Dong}, S., \& {Xie}, J. 2018, \apj, 860,
  101, \dodoi{10.3847/1538-4357/aac6d5}

\end{thebibliography}
